\newcommand{\pa}{\partial}

\newcommand{\been}{\begin{equation}}
\newcommand{\een}{\end{equation}}
\newcommand{\beena}{\begin{eqnarray}}
\newcommand{\eena}{\end{eqnarray}}

\documentclass[12pt]{article}

\usepackage[left=1in,right=1in,top=1in,bottom=1in]{geometry}

\usepackage{amsfonts,amssymb,amsmath,amsthm}
\usepackage{graphicx,color,subfigure,psfrag}

\def\beq#1#2\eeq{\begin{equation}\label{#1}#2\end{equation}}
\def\bal#1#2\eal{\begin{align}\label{#1}#2\end{align}}
\def\bse#1#2\ese{\begin{subequations}\label{#1}#2\end{subequations}}
\def\ba{\begin{aligned}}
\def\ea{\end{aligned}}

\newtheorem{thm}{Theorem}
\def\dd{\operatorname{d}}
\def\C{C}
\def\D{D}

\def\ann#1{#1} \def\wjp#1{#1}

\begin{document} 
\def\singlespacing{\baselineskip=13pt}	\def\doublespacing{\baselineskip=18pt}
\singlespacing

\pagestyle{myheadings}\markright{{\sc Active exterior  cloaking}  ~~~~~~\today}

\title{Source amplitudes  for  active exterior  cloaking }

\author{ A.  N. Norris$^{a}$,  F. A. Amirkulova$^{a}$, W. J. Parnell$^{b}$ \\ ~ \\
(a) Mechanical and Aerospace Engineering,  	Rutgers University, \\ Piscataway NJ 08854-8058, USA \\
(b) School of Mathematics, Alan Turing Building, \\ University of Manchester, Oxford Road, \\ Manchester, M13 9PL, UK}

\maketitle
\begin{abstract}

The  active cloak  comprises a discrete set of multipole sources that destructively interfere with an  incident time harmonic scalar wave to produce zero total field over a finite spatial region.  For a given number of sources and their   positions in two dimensions it is shown  that the  multipole amplitudes  can be expressed  as infinite sums of the coefficients of the incident wave decomposed into regular Bessel functions.
\ann{The field generated by the active sources vanishes in the infinite region exterior to a set of circles defined by the relative positions of the sources.
}
The results  provide a direct solution to the inverse problem of determining the source amplitudes.  They also define a broad class  of non-radiating discrete sources.

\end{abstract}


\section{Introduction}\label{sec1}

Cloaking is intended to make an object undetectable to incident waves.  The approaches proposed  consist mainly of  two quite distinct types  of cloaking, namely passive and active.  Passive cloaking
requires devising a metamaterial that can steer the wave energy around the object regardless of the incident wave.  Our interest here is with active cloaking, specifically in situations where the active sources lie in the  exterior of the region containing the cloaked object.  We call this configuration  {\it active exterior cloaking} in keeping with prior terminology  \cite{Vasquez09}.

Despite the dominant interest in passive cloaking devices, active exterior cloaking has been  investigated quite extensively   \cite{Miller06,Vasquez09,Vasquez09b,Vasquez11a,Vasquez11}.    Miller  \cite{ Miller06}  proposed
creating a cloaking region by measuring   particle motion near the surface of the cloaking zone while simultaneously  exciting appropriate surface sources where each source amplitude depends on the measurements at all sensing points.  As an   active  cloaking method  this approach is  limited because it does not provide a unique relationship between the incident field on the one hand, and the source amplitudes on the other.   A  solution to this problem   was provided  by Vasquez et al.\  \cite{Vasquez09,Vasquez09b} in the context of    active exterior cloaking for the 2D Helmholtz equation.  They used Green's formula and addition theorems for Bessel functions  to formulate an integral equation, which was converted to a linear system of equations for the unknown amplitudes.  Crucially, the integral equation provides the source  amplitudes as linear functions of the incident wave field.  Vasquez et al.\ also showed, by construction, that active cloaking can be realized using as few as three active sources in 2D.
A more explicit form of the linear relation for the source amplitudes as a function of the incident field was developed in \cite{Vasquez11}.  Multipolar sources were used to reproduce Miller's cloak \cite{Miller06}, and  numerical results were compared with SVD solutions of the linearized system \cite{Vasquez09,Vasquez09b}.  The approach of  \cite{Vasquez09,Vasquez09b}  was generalized in  \cite{Vasquez11a} to handle the 3D Helmholtz equation, seeking non resonant frequencies of the cloaked object.  Further analysis and extension of the methods to the quasistatic regime relevant to Laplace's equation can be found in \cite{Onofrei11,Vasquez11b,Milton06b}.   The  active source method of Vasquez et al. has also been adapted  to  create illusion effects so that an object outside the cloaking region  can be made to appear like another object
\cite{Zheng10}.

In this paper  we demonstrate that the  integral representation of  Vasquez et al.\  \cite{Vasquez11} for  the source amplitudes can be reduced to closed-form explicit formulas. This obviates the need to reduce the integral equation of Vasquez et al.\  \cite{Vasquez09,Vasquez09b} to a system of linear equations, which must then be solved numerically, or to evaluate line integrals, as proposed in  \cite{Vasquez11}.  We provide analytical expressions for the source amplitude coefficients for  general incidence as well as plane wave incidence.   The expressions involve no more than sums  of cylinder  functions which can be truncated to achieve any desired accuracy.
\ann{We also prove that the field generated by the active sources vanishes in the infinite region exterior to a set of circles defined by the relative positions of the sources.    The active source field, by construction, cancels the incident field in the cloaked region, which is  defined by the  region interior to the same circular areas.
}
The  analytical results are verified by  calculation of the  farfield and the nearfield amplitudes, which  are shown to vanish when the summation is accurately evaluated.

The non-radiating nature of the active field has relevance to the inverse source problem \cite{Tsitsas12}. Although for this problem, some uniqueness results are available for restricted forms of sources, e.g. ``minimum energy sources'' \cite{Devaney85}, in general the solution to this problem is know to be non-unique \cite{Bleistein77}. Here we develop the solution of the active cloaking problem as a new family of non-radiating sources, with the property that they cancel a given incident field over a finite region.

We begin in \S\ref{sec2} with a statement of the problem, a review of the governing  equations, and a summary of the main results, given in eqs.\ \eqref{-11} and \eqref{-9}.   The basic integral relation of Vasquez et al.\  \cite{Vasquez11} is derived in \S\ref{sec3}, from which the main results are shown to follow.
\ann{
Some example applications of the new formulas are presented in   \S\ref{sec4}.  Some implications of the general results are   discussed in \S\ref{sec5}    and conclusions are given in \S\ref{sec6}.
}

\begin{figure}[htb]	
	\centering
		\includegraphics[width=4.5in]{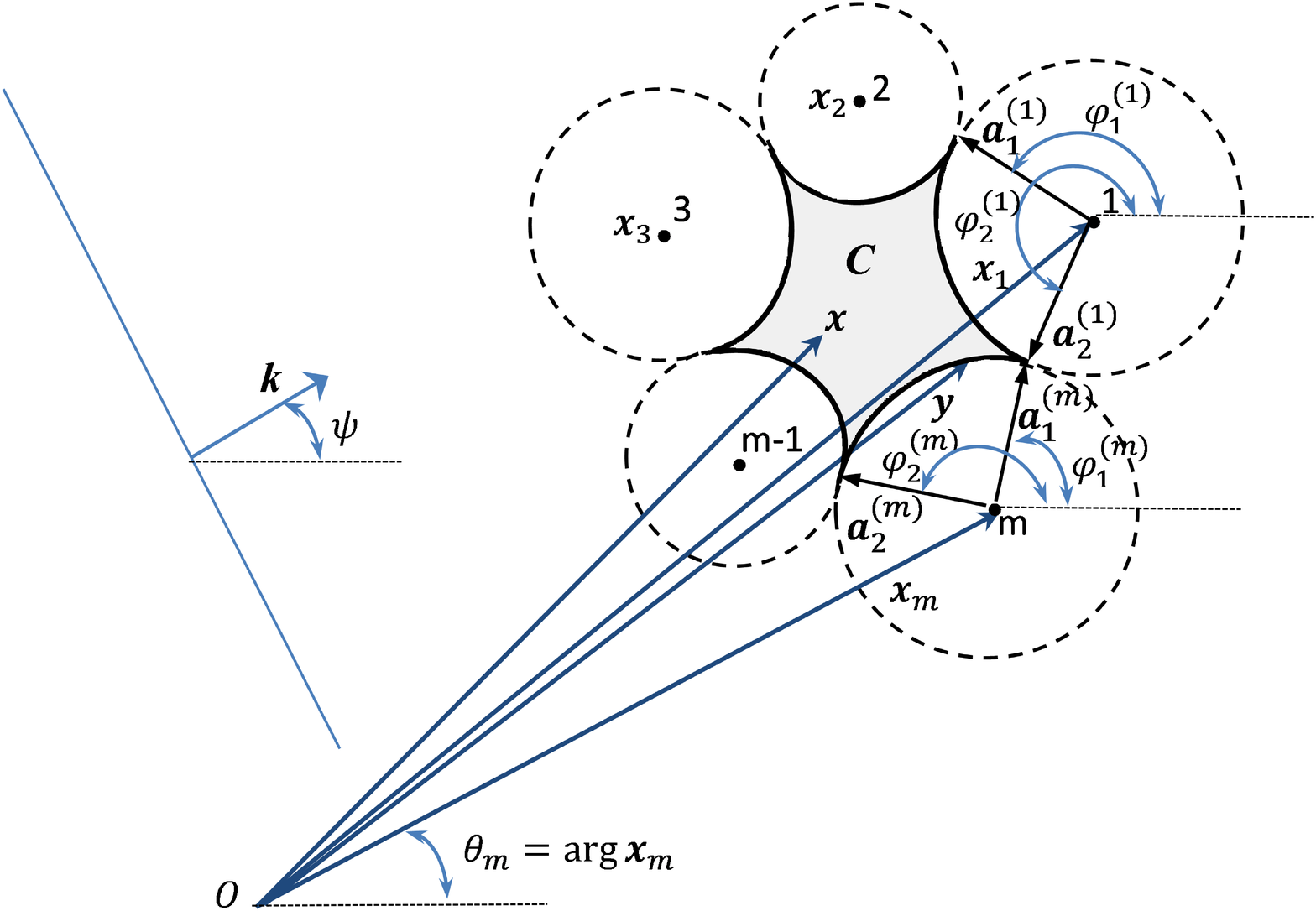}
\caption{Insonification of the  actively cloaked region $\C$ generated by  $M$ active point multipole sources  at ${\bf x}_m$, $m=\overline{1,M}$.   The region $R$ is defined as the interior of
the union of the dashed circular arcs, that is, the combined area comprising  $C$ and the $M$ circular domains.
The incident field in this case is a  plane wave with wave vector $\textbf{k}$
in the direction  $\psi$.
}
\label{fig:fig_1}
\end{figure}

\section{The problem and its solution}\label{sec2}

\subsection{Problem overview}

The active cloaking devices considered here operate in two dimensions, and consist of arrays of   point multipole sources located at positions ${\bf x}_m \in \mathbb{R}^2$, $m=\overline{1,M}$,
see Figure \ref{fig:fig_1}. The active sources  lie in the exterior region with respect to the  cloaked region $\C$
and  this type of cloaking may therefore be  called ``active exterior cloaking" \cite{Vasquez09}.
Objects are undetectable in the cloaked region by virtue of the destructive interference of the sources and  the incident field with the result that  the total wave amplitude  vanishes in the cloaked region $\C$.  The advantages of this type of  cloaking device are:
(i) the cloaked region is  not completely surrounded by a single cloaking device; (ii) only a small  number of active sources are needed; (iii)  the procedure  works for broadband input sources.  \wjp{A disadvantage of the active cloaking approach is that the fields near the ideal sources may become uncontrollably large. Realistically these would be replaced by regions of finite extent and thus their magnitude is reduced.  A further disadvantage of the method is that  the incident field must be known.
However we note that with the approach proposed in this paper, the new expressions require only the expansion of the incident field into entire cylindrical waves, as compared with the line integrals derived in \cite{Vasquez11a} which require knowledge of the incident field and its normal derivative.}

   The shaded region in Figure \ref{fig:fig_1} denotes the  cloaked zone $\C$ generated by $M$ active point multipole sources.  The boundary of $\C$  is the closed concave union of the circular arcs
 $m=\overline{1,M}$, $\{ a_m, \phi^{(m)}_{1}, \phi^{(m)}_{2} \}$  associated with the source at ${\bf x}_m$. In the general case $\{ a_m, \phi^{(m)}_{1}, \phi^{(m)}_{2} \}$ are distinct for different values of $m$.  Note that the wave incidence shown in Figure \ref{fig:fig_1} is a plane wave  although the solution  derived below is for arbitrary incidence.
 The inverse problem to be solved is to find the amplitudes of the active sources as a function of the incident wave, and to prove that the cloaked region is indeed the closed region  $\C$.

\subsection{Basic equations}

We assume time harmonic dependence $e^{-i\omega t}$ which is omitted hereafter and consider the scalar Helmholtz equation in two dimensions. Thus the method proposed here is applicable to any physical situation described as such. For ease of discussion however let us consider the case of acoustics, so that the governing equation for the (time harmonic) pressure $u({\bf x})$ is
\beq{-1}
\nabla^2 u + k^2 u = s ,
\eeq
where $ k = \omega /c$ is the wavenumber, $c$ the acoustic speed, and the term $s$ represents sources.
For a given incident wave  we assume there is an additional field  resulting from the active sources  which exactly cancels the incident wave in some bounded region $\C$.
This additional wave field  is caused by  the $M$ multipole sources located at ${\bf x}_m$, $m=\overline{1,M}$.
The assumed form of the  total field $u$, the  incident wave $u_i$, and the
active source  field $u_d$ are, respectively
\begin{subequations}\label{-11}
\bal{-110}
u & = u_i + u_d ,
\\
u_i &=
 \sum\limits_{n=-\infty}^\infty
A_n U_n^{\,+}({\bf x} ),
\label{-11a}
\\
u_d^{} &=
\sum\limits_{m=1}^M\sum\limits_{n=-\infty}^\infty
b_{m,n} V_n^{\,+}({\bf x} - {\bf x}_m), \label{-11b}
\eal
\end{subequations}
where  the wave functions $U_n^{\,\pm}({\bf x} )$  and   $V_n^{\,\pm}({\bf x} )$  are defined by
\beq{12-}
U_n^{\,\pm}({\bf x} ) = J_n (k|{\bf x}|) e^{\pm i n \arg {\bf x} },
\qquad
V_n^{\,\pm}({\bf x} ) = H_n^{(1)} (k|{\bf x}|) e^{\pm i n \arg {\bf x} }.
\eeq
Here $\arg {\bf x} \in [ 0, 2 \pi)$ and $\arg {(- \,{\bf x})} = \arg {\bf x} \pm \pi  \in [ 0, 2 \pi)$.
Define the derivative functions ${U_n^{\,\pm}}'({\bf a})$ as
\beq{8-}
{U_n^{\,\pm}}'({\bf a} ) = J_n'(k a)e^{\pm i n \arg {\bf a} }.
\eeq
In the following we write $U_0$ and $V_0$, with obvious meaning.
Note that the  functions
 $U_n^{\,\pm}({\bf x} )$  and   $V_n^{\,\pm}({\bf x} )$   possess the properties
\beq{13-}
U_n^{\,\pm}(-  {\bf x} ) = (-1)^n U_n^{\,\pm}({\bf x} ),
\qquad
V_n^{\,\pm}(- {\bf x} ) = (-1)^n V_n^{\,\pm}({\bf x} ).
\eeq
The active source field  $u_d^{}$ in \eqref{-11b} is of the same form as considered by Vasquez et al.\
\cite[eq.\ (5)]{Vasquez09b}.  
The three dimensional analog is given in \cite[eq.\ (40)]{Vasquez11a}.
The  coefficients $A_n$, which define the incident field, include as a special case plane wave incidence in the direction $\psi$  $(A_n = i^n e^{-in\psi} )$.

The active cloaking problem is now to find (i) the coefficients $b_{m,n}$ such that the total field $u$ vanishes inside some compact region $\C$, and (ii) to define the region $\C$.

\subsection{Summary of main results}
The principal  results can be summarized in two theorems.  The first provides necessary and sufficient conditions on the source amplitudes $b_{m,n}$ in order to ensure cloaking in the region $\C$ and a non-radiating source field $u_d$. The second provides the explicit expressions for the active source amplitudes.

\begin{thm}\label{thm1}
Necessary and sufficient conditions on the active source coefficients  $b_{m,l} $ in order to ensure zero total field ($u_i+u_d=0$) inside $\C$ and no radiated field ($u_d\rightarrow 0$ in the far field) are
\beq{14}
\forall \, n\in \mathbb{Z}:  \quad
\sum\limits_{m=1}^M\sum\limits_{l=-\infty}^\infty
b_{m,l} \,
\times
\begin{cases}
  U_{n-l}^{-}({\bf x}_m)  &= 0,
	\\ &  \\
	   V_{n-l}^{-}({\bf x}_m)  &= -A_n .
\end{cases}	
	\eeq
\end{thm}

These identities provide a useful means to quantify error in active cloaking as will be seen later on. We now state the explicit form for the source amplitudes, together with the shape of the cloaked region $\C$ and the region in which the source field vanishes.

\begin{thm}\label{thm2}
Given $M$ active sources located at ${\bf x}_m$, $m=\overline{1,M}$, the required active source amplitude coefficients
for the general incidence \eqref{-11a} are
\begin{subequations}\label{-9}
\bal{-9a}
b_{m,l} & = \sum\limits_{n=-\infty}^\infty
b_{m,ln} A_n  \quad \text{where}
\\
b_{m,ln}   & =
\frac  {k a_m}4
\sum\limits_{p=-\infty}^\infty U_{n+p}^{\,+}( {\bf x}_m )
 \frac {(-1)^{p} } {l+p}
\big[
J_p(k a_m)J_l'(k a_m) - J_p'(k a_m)J_l(k a_m)
\big] 	
		\nonumber \\ & \, \qquad \qquad \times \,
	\big[
e^{-i ( l+p) \phi^{(m)}_{2} } - e^{-i ( l+p) \phi^{(m)}_{1} }
\big].
\label{-9b}
\eal
\end{subequations}
This ensures cloaking (zero total field) in the region $\C$ which is the closed and bounded domain formed by taking its boundary as the closed concave union of the circular arcs defined by $\{ a_m, \phi^{(m)}_{1}, \phi^{(m)}_{2} \}$ and denoted as $\partial C_m$, see Figure \ref{fig:fig_1}.
\ann{These coefficients also ensure that the  radiated field from $u_d$ is identically zero in the region exterior to all of the circles centered at the source points:
\beq{2=1}
u_d ({\bf x}) = 0
\ \ \textnormal{for}\
 {\bf x} \in \mathbb{R}^2 /R,
\quad R  \equiv   \C  \bigcup \{ {\bf x}:\,  |{\bf x}-{\bf x}_m| \le a_m,\, m=\overline{1,M} \}   .
\eeq
This is the exterior to the union of the dashed circular arcs in Figure \ref{fig:fig_1}.
}

\end{thm}

 An alternative and more concise formulation of eq.\ \eqref{-9b} is obtained  using the notation of eq.\ \eqref{12-} with
{${\bf a}_i^{(m)} \equiv  a_m \hat{\bf e}(\phi_i^m), (i=1,\,2)$},
\beq{-7}
b_{m,ln} =
\frac 14 {k a_m}
\sum\limits_{p=-\infty}^\infty U_{n+p}^{\,+}( {\bf x}_m )
 \frac {(-1)^{p} } {l+p}
 \left.
   \big[
   {
   {U_p^{-}}({\bf {a}}) {U_l^{-}}'({\bf {a}})- {U_p^{-}}'({\bf {a}}) {U_l^{-}}({\bf {a}})
   }\big] \right|_{{\bf a}_1^{(m)}}^{{\bf a}_2^{(m)}}
 \eeq
where $\hat{\bf e}(\phi_i^m)$ is a unit vector subtended at angle $\phi_i^m$, as illustrated in Figure \ref{fig:fig_1}.

An important case for which the summation in \eqref{-9a} can be simplified is  plane wave incidence.  Assuming the incident field is  a unit amplitude plane wave in direction $\psi$, $u_i = u_\psi$ defined by   $A_n = i^n e^{-in\psi} $,  results in
\beq{6-}
b_{m,l}  =
u_\psi({\bf x}_m) \frac {k a_m}4
\sum\limits_{p=-\infty}^\infty
 \frac {i^{p} e^{i   p  \psi}} {l+p}	
 \left.
  \big[ {
   {U_p^{-}}({\bf {a}}) {U_l^{-}}'({\bf {a}})- {U_p^{-}}'({\bf {a}}) {U_l^{-}}({\bf {a}})
   }
    \big]
  \right|_{{\bf a}_1^{(m)}}^{{\bf a}_2^{(m)}} ,
			\ \
			\begin{matrix}\text{plane wave} \\ \text{incidence}. \end{matrix}
\eeq
The form of the coefficients $b_{m,l}$ is discussed further below.  
Note that the term in \eqref{-9b}, \eqref{-7}  and in \eqref{6-},  corresponding to  $p+l=0$ is zero, which follows from l'H\^opital's rule, or otherwise.


Theorems \ref{thm1} and \ref{thm2} are proved in the next section.

\section{Proofs of Theorems 1 and 2} \label{sec3}

\subsection{Theorem 1: Necessary and sufficient conditions on the source amplitudes}

We first prove  the  constraints  on the source coefficients  $b_{m,l} $
given by   Theorem  \ref{thm1}, and at the same time show that they may be interpreted in terms of the near- and far-field of the active sources.
To this end, we express  $u_d^{}$ in two different  forms
using the generalized Graf addition theorem \cite[eq.\ (9.1.79)]{Abramowitz74},
\beq{-2}
V_l^{\,+}({\bf x} - {\bf y})
= \sum\limits_{n=-\infty}^\infty
\begin{cases}
V_n^{\,+}({\bf x} )\,
 U_{n-l}^{\,-}({\bf y}) ,   & |{\bf x}|>|{\bf y}|,
\\
U_n^{\,+}({\bf x} )\,
 V_{n-l}^{\,-}({\bf y})   , & |{\bf x}|<|{\bf y}|.
\end{cases}
\eeq

Let us first consider the radiated field, assuming that $u_d$ does not radiate energy into the far field. The first of \eqref{-2}, for $|{\bf x}|>|{\bf y}|$,  allows us to  rewrite $u_d^{}$ as a sum of multipoles at the origin:
\beq{166}
u_d^{} =
\sum\limits_{n=-\infty}^\infty F_n V_n^{\,+}({\bf x} )
\ \ \text{ for }
\ \  |{\bf x}|> \text{max}(|{\bf x}_m|+a_m),
\eeq
where
\beq{444}
 F_n =
\sum\limits_{m=1}^M\sum\limits_{l=-\infty}^\infty
b_{m,l}   U_{n-l}^{\,-}({\bf x}_m).
\eeq
Define    the farfield amplitude function $f(\theta ) $,
  $\theta = \arg \hat {\bf x}$, such that
\beq{-33}
u_d^{}({\bf x}) = f(\theta ) \,
\frac{ e^{ i k |  {\bf x}|  } }{ (k |  {\bf x}|)^{1/2} }
+ \text{O}\big( (k |  {\bf x}|)^{-3/2}\big), \quad
|{\bf x}|\rightarrow \infty .
\eeq
The farfield amplitude function  follows from the asymptotic form of the Hankel functions  as
\beq{12}
f(\theta ) = \sum\limits_{n=-\infty}^\infty f_n e^{i n \theta } ,
\quad f_n = \Big( \frac 2{ \pi}\Big)^{1/2} \, i^{-(n +\frac 12 )} \,  F_n .
\eeq
A measure of the  nondimensional total power radiated by the sources  is  given by the
non-negative far-field flux parameter
\beq{44}
\sigma_{r} = \int_0^{2\pi} \dd \theta |f(\theta)|^2
= 4  \sum\limits_{n=-\infty}^\infty |F_n|^2 .
\eeq
Since $u_d$ does not radiate energy into the far field, the active sources must vanish, so that $F_n = 0\, \forall n$. Imposing this in \eqref{444} ensures the necessity of \eqref{14}$_1$.  The sufficiency of \eqref{14}$_1$ is seen immediately by substituting \eqref{14}$_1$ into \eqref{444} and \eqref{166} which gives $u_d=0$.


Now let us consider the near-field inside the cloaked region $\C$ where we assume that the cloaked region contains the origin and the total field is zero inside $\C$, i.e.\ $u_i+u_d=0$. Using the second identity in \eqref{-2}, the active source field $u_d^{}$ can be expressed in a form that is valid in the neighborhood of the origin (assuming $|{\bf x}_m|>a_m\, \forall m$),
\beq{167}
u_d^{} =
\sum\limits_{n=-\infty}^\infty E_n U_n^{\,+}({\bf x} )
\ \ \text{ for }
\ \   |{\bf x}|< \text{min}(|{\bf x}_m|-a_m),
\eeq
where
\beq{033}
E_n =
\sum\limits_{m=1}^M\sum\limits_{l=-\infty}^\infty
b_{m,l}     V_{n-l}^{\,-}({\bf x}_m).
\eeq
The total field vanishing in some
neighbourhood of the origin thus implies that $E_n+A_n $ vanishes for  every value of $n$. This gives rise to the necessary condition \eqref{14}$_2$. Sufficiency is once again immediate by assuming the form \eqref{14}$_2$ and back-substituting into the forms of $u_d$ and $u_i$ above.

Further implications of this result are explored after we   complete the proof of
Theorem  \ref{thm2}.

\subsection{Theorem 2: Explicit forms for the active source amplitudes} 
\label{233}

\begin{figure}[htb]	
	\centering
\psfrag{C}{$\D$}
\psfrag{x1}{${\bf x}_1$}
\psfrag{x2}{${\bf x}_2$}
\psfrag{x3}{${\bf x}_3$}
\psfrag{x4}{${\bf x}_4$}
\psfrag{C1}{$\pa D_1$}
\psfrag{C2}{$\pa D_2$}
\psfrag{C3}{$\pa D_3$}
\psfrag{C4}{$\pa D_4$}
		\includegraphics[width=2.4in]{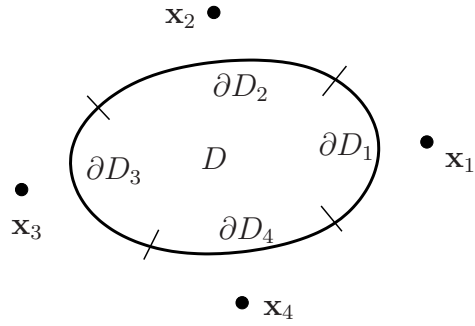}    
\caption{
A configuration of $M=4$ sources, and a region $D$ in which the integral identity
\eqref{-0} holds.
}
\label{fig2}
\end{figure}

\wjp{The Green's function $g({\bf x}, {\bf x}') $ is defined as the solution of \eqref{-1} for source
$s = \delta ( {\bf x}- {\bf x}')$, i.e. $g({\bf x}, {\bf x}')  = -\frac i4 V_0 ({\bf x}-{\bf x}')$.
Consider a region $\D$ such as that depicted in Figure \ref{fig2}, chosen so that it does not contain any sources. We will determine the explicit form for the active source amplitudes together with the form of $\D$ that ensures cloaking. The latter, already introduced as $\C$, is the region depicted in Figure \ref{fig:fig_1}.}

\begin{figure}[htb]	
	\centering
\psfrag{x1}{$\bf{x}_1$}
\psfrag{x2}{$\bf{x}_2$}
\psfrag{x3}{$\bf{x}_3$}
\psfrag{c1}{$\partial \C_1$}
\psfrag{c2}{$\partial \C_2$}
\psfrag{c3}{$\partial \C_3$}
		\includegraphics[width=2.3in]{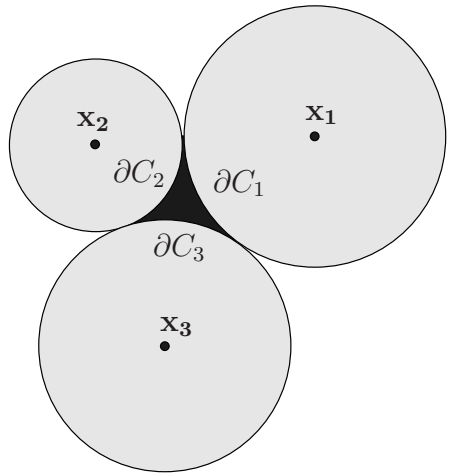}    
\caption{
The  integration curve $\pa C$   split into $M=3$ portions $\partial C_m$ appropriate for the integral
representation \eqref{minus} of the active source field.   The cloaked central (black) region, is bounded by  $\partial C_m$, $m=1,2,3$.
  }
\label{fig3}
\end{figure}

\wjp{
By assumption, both $u_i$ and $u_d$ satisfy the homogeneous Helmholtz equation in $\D$  (eq.\ \eqref{-1} with $s=0$ $\forall\, {\bf x}\in \D$), and therefore
 \beq{-0}
\int_{\partial D} \dd S ({\bf y})  \big[
v({\bf y}) \partial_n  g ( {\bf y} , {\bf x})
- g ( {\bf y} , {\bf x})
\partial_n v({\bf y})
\big]
 = v ({\bf x}) , \ \ v=\{u_i, u_d\}, \ \ {\bf x}\in \D.
\eeq
where $\pa D$ is the boundary of $\D$ depicted in Figure \ref{fig2} as the union of the arcs $\pa D_m, m=\overline{1,M}$ and it is traversed counter-clockwise. We wish to determine the cloaked region $\C\subset\D$ which is defined by its property that
the total field $u_i + u_d$ vanishes inside $\C$, so that
 \beq{-}
u_d ({\bf x})   = -u_i ({\bf x})= \frac i 4
\int_{\partial C} \dd S ({\bf y})  \big[
u_i({\bf y}) \partial_n  V_0  ( {\bf y} - {\bf x})
-V_0   (  {\bf y} - {\bf x})
\partial_n u_i({\bf y})
\big], \ \  {\bf x} \in \C .
\eeq
Given that the boundary of $C$ is split up, as for $D$ into segments $\pa C_m, m=\overline{1,M}$, we can use \eqref{-2}$_1$, in order to write, for some $\bf{x}_0$
\bal{-49}
V_0({\bf y}-{\bf x}) 
= V_0({\bf x}-{\bf x}_0-({\bf y}-{\bf x}_0)) 
= \sum_{n=-\infty}^{\infty}V_n^+({\bf x}-{\bf x}_0)U_n^-({\bf y}-{\bf x}_0)
\eal
which holds for $|{\bf x}-{\bf x}_0|>|{\bf y}-{\bf x}_0|$. Do this for each of the contours choosing ${\bf x}_0={\bf x}_m$ on each $\pa C_m$, , so that
\beq{minus}
u_d({\bf x}) = -\frac{i}{4} \sum_{m=1}^M
\sum_{n=-\infty}^{\infty} V_n^+({\bf x}-{\bf x}_m)
\int_{\pa C_m} \dd S_m\left(u_i({\bf y})\partial_n U^-_n({\bf y}-{\bf x}_m) - U_n^-({\bf y}-{\bf x}_m)\partial_n u_i({\bf y})\right)
\eeq
where we require $|{\bf x}-{\bf x}_m|>|{\bf y}-{\bf x}_m|$ on each contour $\pa C_m$ (recall that the integral is being considered for ${\bf x} \in \C$). The minus sign in \eqref{minus} arises since upon expanding about the point ${\bf x}_m$, the counter-clockwise orientation with respect to the centre ${\bf x}_m$ is opposite to the counter-clockwise traversal of $\pa C$ with respect to some origin inside $\C$. Note that for this to hold simultaneously for all $m$ the contours $\pa C_m$ must be circular arcs as depicted in Figure \ref{fig3}.
Therefore we have proved that $\C$ is the region with boundary as the closed concave union of the circular arcs defined by $\{ a_m, \phi^{(m)}_{1}, \phi^{(m)}_{2} \}$ and denoted as $\partial C_m$, see Figure \ref{fig:fig_1}.
Finally, using the form for $u_d$ given in \eqref{-11b}, we find that
\beq{2}
b_{m,n}  = -\frac i 4
\int_{\pa C_m} \dd S_m \Big[
u_i({\bf y}) \partial_n U_n^- ( {\bf y} - {\bf x}_m)
-U_n^- ( {\bf y} - {\bf x}_m)
\partial_n u_i({\bf y})
\Big] .
\eeq
}
This agrees with   \cite[Eq.\ (8)]{Vasquez11} apart from a  factor $i/4$  missing there.  Equation \eqref{2} provides a direct method for calculating the multipole source amplitudes, as  has been demonstrated  numerically for different source configurations    \cite{Vasquez11}.   The result is not optimal, however, as it requires evaluation of a line integral, which can be
computationally  time consuming.

\wjp{
The explicit formula for the source amplitudes follows from eq.\ \eqref{2} by introducing the forms for the functions $U_n^-$ as follows
\beq{3}
b_{m,l}   = {-}\frac i 4  k a_m
\int_{\phi_1^{(m)}}^{\phi_2^{(m)}} \dd \phi \, e^{-i l \phi } \big[
u_i({\bf y})  J_l' ( ka_m)
-J_l( ka_m)  k^{-1}
\partial_n u_i({\bf y})
\big]
.
\eeq
We see that the cloaked region $\C$ is indeed the subdomain of $D$ in which Graf's theorem can be simultaneously invoked for all of the $M$ active sources.
}

Consider plane wave incidence in the direction of the unit vector $\hat{\bf e}(\psi)$,
$u_i = u_{\psi}({\bf x})$ where
\beq{4}
 u_{\psi}({\bf x}) = e^{i k  \hat{\bf e}(\psi) \cdot {\bf x}}
 \quad (A_n = i^n e^{-in\psi)} )
.
\eeq
Then \eqref{3} becomes, with  $\alpha_m \equiv ka_m$,
\bal{5}
b_{m,l}   &= {\frac{\alpha_m}{4i}}  u_\psi({\bf x}_m)
\int_{\phi_1^{(m)}}^{\phi_2^{(m)}} \dd \phi \, e^{- il \phi }  \big[
  J_l' ( \alpha_m)
-i {\bf n}(\phi)  \cdot
\hat{\bf e}(\psi)J_l( \alpha_m)
 \big]
 u_\psi({\bf y}-{\bf x}_m)
\nonumber \\
&=  {\frac{\alpha_m}{4i}}  u_\psi({\bf x}_m)
\int_{\phi_1^{(m)}}^{\phi_2^{(m)}} \dd \phi \big[
  J_l' ( \alpha_m)
-i  \cos(\phi-\psi)  J_l( \alpha_m)
 \big]
e^{i [ \alpha_j \cos(\phi-\psi) - l \phi ]}
\nonumber \\
&= {\frac{\alpha_m}{4i}}  u_\psi({\bf x}_m) \, e^{- il \psi }
  \big[ J_l' ( \alpha_m)  G( \alpha_m) - J_l ( \alpha_m)  G '( \alpha_m)
 \big]
,
\eal
where the function $G$ is defined as
\beq{6}
G(\alpha) =
\int_{\phi_1^{(m)}-\psi }^{\phi_2^{(m)}-\psi } \dd \phi \,
e^{i ( \alpha \cos \phi-   l \phi )}
=
 \sum\limits_{n=-\infty}^\infty
J_n(\alpha) \, i^{n}
\int_{\phi_1^{(m)}-\psi }^{\phi_2^{(m)}-\psi } \dd \phi \,
e^{-i ( n+   l) \phi } .
\eeq
The identity $
e^{i x \sin\theta} = \sum_{n=-\infty}^\infty
J_n(x) e^{i n\theta}$
has been used in simplifying  the form of  $G(\alpha)$.
Performing the integration in \eqref{6}, we arrive at an  explicit expression for the amplitude coefficients
\beq{9}
b_{m,l}   =
u_\psi({\bf x}_m) \frac {\alpha_m}4
\sum\limits_{p=-\infty}^\infty
{
\big[
J_p(\alpha_m)J_l'(\alpha_m) - J_p'(\alpha_m)J_l(\alpha_m)
\big]
    \frac {i^{p} e^{i   p  \psi}} {p+l}	
	\big[
e^{-i (p+l) \phi^{(m)}_{2} } - e^{-i (p+l) \phi^{(m)}_{1} }
\big]
}
    .
\eeq
Now consider  the incident field
\beq{43}
 \frac{i^{-n}} {2\pi}\int_0^{2\pi} \dd \psi \, u_{\psi}({\bf x}) e^{in\psi}
= U_n^{\,+}( {\bf x} )
 \quad (A_p = \delta_{np}).
\eeq
It follows from integration of \eqref{9} that the general form of the amplitude coefficients
for the general incidence \eqref{-11a} is given by \eqref{-9b}.

\ann{
Finally, we turn to the question of where the active source field vanishes, noting  that    the integral \eqref{-}  vanishes identically for field positions outside $\C$ \cite{Colton}
 \beq{-3}
\frac i 4
\int_{\partial C} \dd S ({\bf y})  \big[
u_i({\bf y}) \partial_n  V_0  ( {\bf y} - {\bf x})
-V_0   (  {\bf y} - {\bf x})
\partial_n u_i({\bf y})
\big] = 0 , \ \  {\bf x} \in \mathbb{R}^2/\C .
\eeq
How does this relate to the  source field $u_d ({\bf x})$?  In the course of the  derivation of the coefficients $b_{m,n}$  the field $u_d ({\bf x})$ was expressed in the form \eqref{minus} for
${\bf x} \in\C$.  The latter restriction on ${\bf x}$ can be removed since it is clear that eq.\ \eqref{minus}
{\it defines}   $u_d ({\bf x})$ for all ${\bf x}$.  This is evident from the definition \eqref{-11b} and from the identity \eqref{2} for $b_{m,n}$.   Equation \eqref{-3} therefore implies that  $u_d ({\bf x})  $ vanishes
at  all positions outside the cloaked region for which the representation \eqref{minus} holds, i.e.\
$\{ {\bf x} \not\in \C:\    |{\bf x}-{\bf x}_m|>|{\bf y}-{\bf x}_m|,\,  {\bf y}\in \pa C_m, \, m=\overline{1,M} \}$.
This is precisely the region $R$ defined in \eqref{2=1}, equal to, for instance, the exterior to the colored regions in Figure \ref{fig3}.
}


This completes the proof of Theorem \ref{thm2}.


\begin{figure}[htb]	
	\centering
		\includegraphics[width=3.5in]{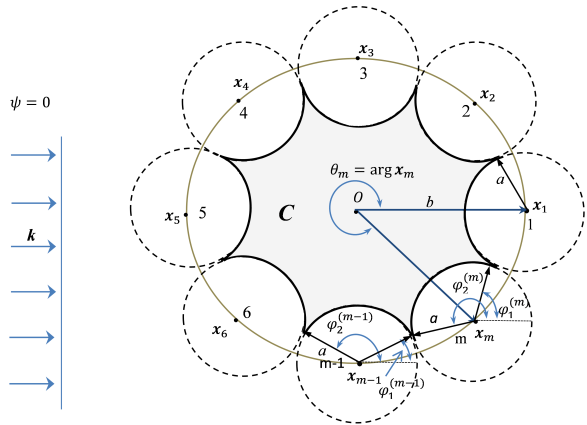}  
\caption{Plane wave insonification of the cloaking region $\C$ generated by  $M = 8$ active sources.}
\label{fig:fig_4}
\end{figure}

\section{Numerical examples} \label{sec4}

\subsection{Active source configuration} 			\label{Config}

We illustrate the  results for plane wave incidence on configurations of the type shown in Fig.\ \ref{fig:fig_4}.  The  $M$ sources are symmetrically located on a circle,  with
\beq{-23}
a_m = a, \quad
|{\bf x}_m|=b,\quad
\theta_m 
= (m-1)\theta_0
\quad m=\overline{1,  M}, \quad \text{where}\ \ \theta_0 = 2\pi /M ,
\eeq
and by necessity,  $
a\ge   b \sin\frac{\pi} M $.
The  circular arcs, which all have the same angular extent, are then defined by
\beq{122}
\phi^{(m)}_{1,2} = \pi + \theta_m \mp
\bigg|
\sin^{-1} \left( \frac ba \sin \frac{\pi} M
\right) - \frac{\pi} M  \bigg|,
\quad m=\overline{1,  M}.
\eeq
We take $ a =   b \sin\frac{\pi} M $  in all examples considered.
Note that the cloaked region $\C$ can be formed by a minimum of 3 sources.  A  configuration with $M=8$ sources  is shown in Figure \ref{fig:fig_4}.   All calculations were performed for plane wave incidence on
 configurations of the type shown in Figure \ref{fig:fig_4} with  varying  numbers  of   sources,
 $M\ge 3$.

\begin{figure}[h!]	
	\centering
		\includegraphics[width=5.0in]{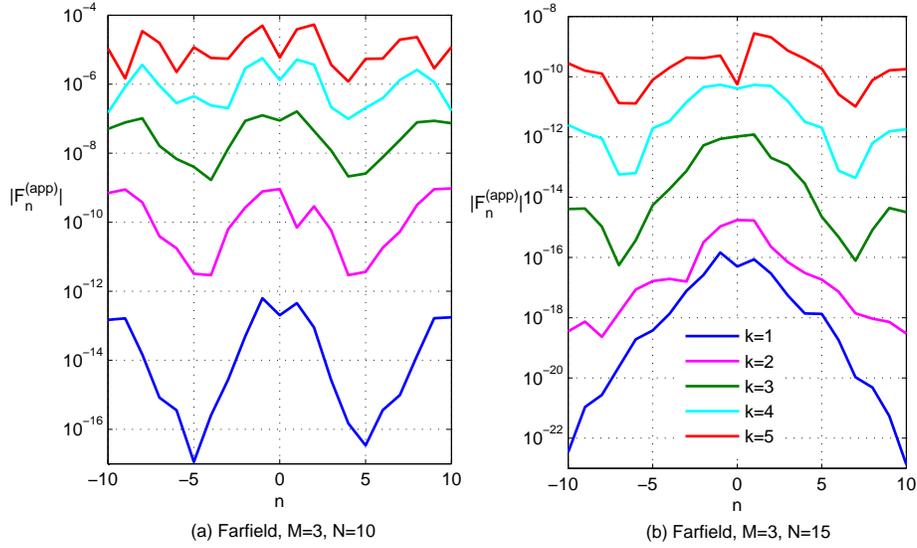}    
\caption{The farfield radiation  amplitudes $|F_n^{(app)}|$ of eq.\ \eqref{122} for different orders of Bessel functions $n = \overline{-10,10}$,  wavenumbers $k=\overline{1,5}$
and truncation values $N=10$ and $15$.   The configuration is
$M=3$ multipole sources  located at the distance $b = 1$ from the origin,   with angle of incidence $\psi=7^\circ$.}
\label{fig_5}
\end{figure}

\begin{figure}[h!]	
	\centering
		\includegraphics[width=5.0in]{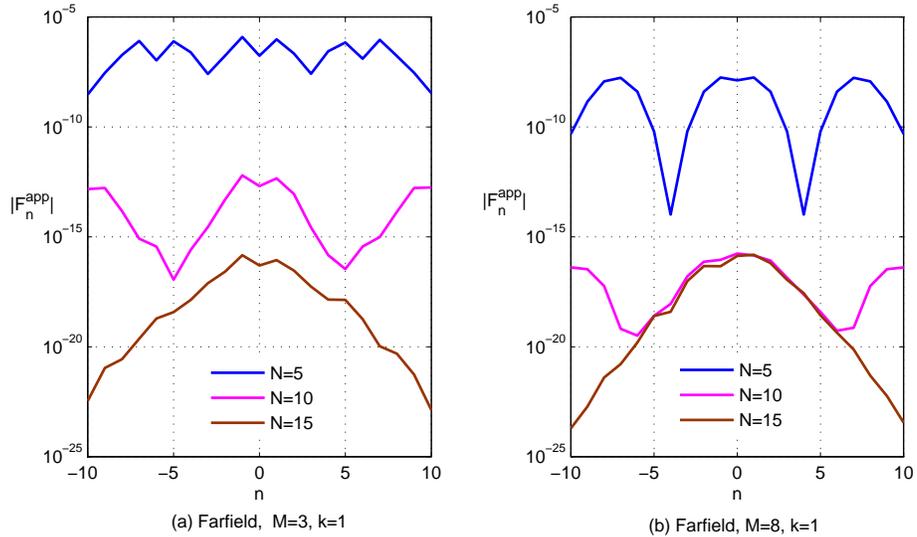}    
\caption{ Dependence of the farfield amplitudes $|F_n^{(app)}|$ on the order $n$ of Bessel functions for different values of  $N$ in  \eqref{14X} ($N=5,10,15$)  and for  different numbers of active sources: (a) $M=3$, and (b) $M=8$.  The  incident wavenumber is $k=1$.
}
\label{fig_6}
\end{figure}

\subsection{Near and farfield amplitudes} \label{sec42}

The efficiency of the cloaked region is  assessed by examining the   farfield
and nearfield as functions of various parameters.  If all terms in
the  infinite sums in eqs.\ \eqref{444} and \eqref{033} are available then the farfield is identically zero and the nearfield exactly cancels the incident wave, by Theorems \ref{thm1} and \ref{thm2}.  We therefore consider
truncated versions of the infinite sums so that  the farfield and nearfield coefficients,  $F_n$ and $E_n$ of eqs.\ \eqref{444} and \eqref{033} respectively, are approximated as
\beq{14X}
\left.
\begin{matrix}
F_n^{(app)}
 \\
E_n^{(app)}
\end{matrix}
\right\}
=\sum\limits_{m=1}^M\sum\limits_{l=-N}^N
b_{m,l} \,	
\times
\begin{cases}
V_{n-l}^{\,-}({\bf x}_m),   
\\
U_{n-l}^{\,-}({\bf x}_m)  ,
 \end{cases}
 	\quad \forall \, n\in \mathbb{Z} .
\eeq
In the limit of $N\to \infty$ exact cloaking is achieved.  Restricting the summation to finite values of $N$ is equivalent to limiting the order of the active multipole sources.
The behavior of the approximate coefficients $F_n^{(app)}$ and $E_n^{(app)}$
has implications on  the accuracy of the cloak regardless of the type of  object to be cloaked.
Thus, the farfield coefficients determine the radiated field everywhere outside the cloak, and must necessarily be small regardless of whether or not an object is being cloaked.  Similarly, the total field in the cloaked region $\C$ must be small in order to achieve cloaking.  The two conditions correspond to   $F_n^{(app)}$ and $E_n^{(app)}+A_n$ having small values.     The  examples in this subsection examine the sensitivity of these quantities.  The sources are located at $b=1$ with   plane waves incident at  $\psi=17^\circ$.

 The farfield amplitude coefficients $|F_n^{(app)}|$, $n=\overline{-10,10}$  are depicted in Figures \ref{fig_5} and  \ref{fig_6} for different values of the wavenumber $k$, the number of sources $M$, and  the number of terms in summation \eqref {14X}, $N$.   It is clear from these two figures that the error in the farfield coefficients decreases  (i) as $N$  increases, (ii) as $M$ increases, and
 (iii) as $k$ decreases.    The convergence is particularly fast as a function of $N$.  For instance, at $k=1$ the farfield coefficients are uniformly less than $10^{-6}$ for all $M\ge 3$ if $N\ge 5$.  Much smaller values ($10^{-15}$ or less)  for  $|F_n^{(app)}|$ are easily achieved for moderate values of $N$, e.g. $N=10$.

\begin{figure}[htb]	
	\centering
		\includegraphics[width=5.0in]{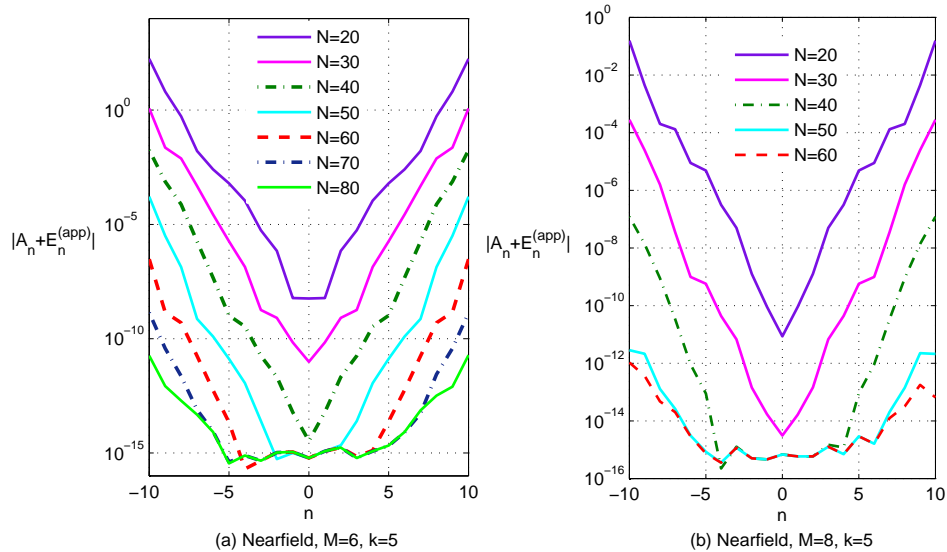}    
\caption{Variation of  the nearfield amplitude coefficients $|A_n+E_n^{(app)}|$ for different values of the truncation size  $N$ in eq.\  \eqref{14X},  generated by $M=6$  active sources in  (a) and $M=8$ sources in  (b).  In all cases   $k=5$.}
\label{fig_7}
\end{figure}

\begin{figure}[htb]	
	\centering
		\includegraphics[width=5.0in]{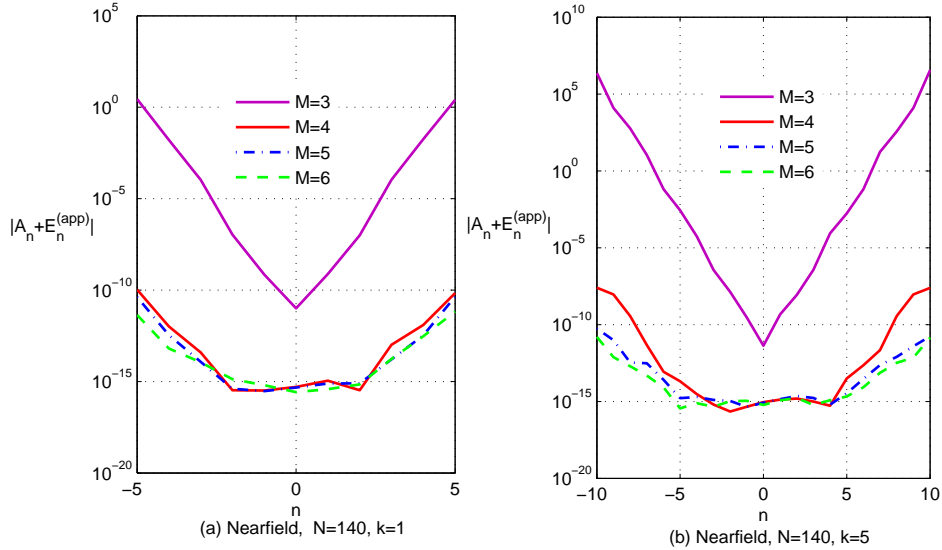}    
\caption{Dependence of the nearfield amplitude $|A_n+E_n^{(app)}|$ on  the number of multipole sources $M$ $(M=3,4,5,6)$ at wavenumber $k=1$ in  (a) and $k=5$ in  (b).   }
\label{fig_8}
\end{figure}

\begin{figure}[h!]	
	\centering
		\includegraphics[width=5.0in]{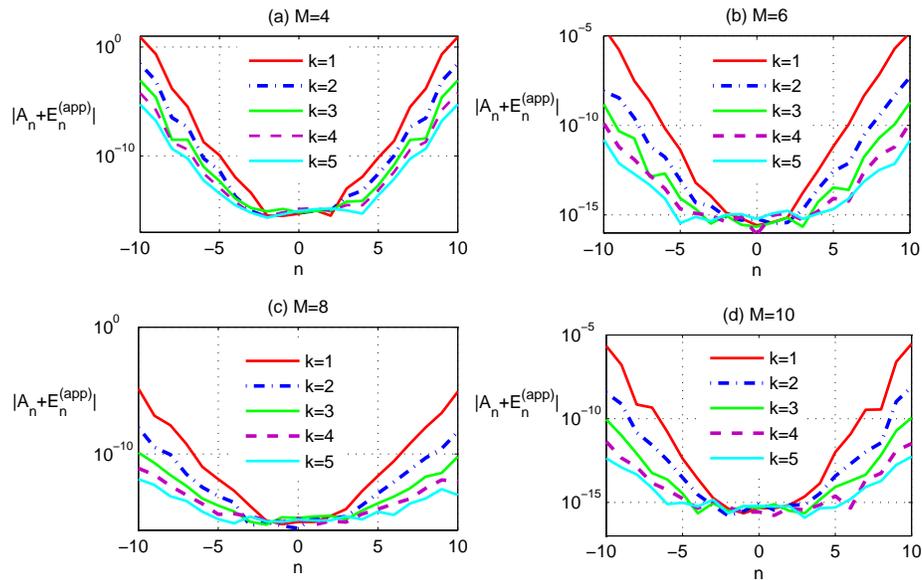}    
\caption{Variation of the nearfield amplitude coefficients with number of active sources
 $(M=4,6,8,10)$ and with wavenumber $(k=\overline{1,5})$.  In all cases $N=130$.  }
\label{fig_9}
\end{figure}

The nearfield amplitude coefficients $|A_n+E_n^{(app)}|$ are shown in Figures \ref{fig_7} to  \ref{fig_9}.  In contrast with the farfield case   relatively large values of the truncation size $N$ are required to obtain small nearfield coefficients.  Figure \ref{fig_7} shows that $N$ on the order of $100$ or more is required to achieve accuracy comparable to the  farfield coefficients.  However, unlike the farfield amplitudes, it is found that the nearfield coefficients generally increase in magnitude with $|n|$, the order of the Bessel functions.  The relatively large values of $|A_n+E_n^{(app)}|$ and their increase with the order $|n|$ does not
necessarily mean that the total field in the nearfield is divergent.  For instance, the top curve in   Figure \ref{fig_7}(a) indicates
$|A_{10}+E_{10}^{(app)}| =$O$(10^2)$, but this value multiplies $J_{10}(kr)$, and, for instance,
$|J_{10}(kr)| < 2\times 10^{-3}$ within $\C$.  In other words, the increasing   values of $|A_n+E_n^{(app)}|$ with $n$ can be balanced by the fact that $J_n(kr) = \frac 1{n!}(\frac{kr}2)^n +\ldots$ for small $kr$.

Figure \ref{fig_8} shows the dependence of the nearfield coefficients on the number of sources.
The case of the minimum number of sources, $M=3$, appears to be strikingly different from others
$(M\ge 4)$.  As Figure \ref{fig_8} indicates,  adding one more source and taking $M=4$ reduces the error from $10^0$ to $10^{-10}$ for $k=1, n=\pm5$ and from $10^{-2}$ to $10^{-14}$ for $k=5, n=\pm5$.  Generally, as with the farfield coefficients,   increasing the number of sources improves the accuracy of the nearfield amplitudes $|A_n+E_n^{(app)}|$.

Finally, Figure \ref{fig_9} shows the nearfield dependence on the  wavenumber, $k=\overline{1,5}$.   The accuracy actually improves with increasing $k$, unlike the farfield case.  However, it should be borne in mind that the nearfield coefficients multiply the terms $J_n (kr)$, which increase in magnitude with $k$ for fixed $r$.

The numerical results in Figures \ref{fig_5} through \ref{fig_9} show  that greater accuracy is achieved using more sources, which is not unexpected.  For the  case of $M=3$, the minimum number required, the nearfield coefficients could be large enough to significantly diminish the cloaking effect.   This suggests taking $M=4$ might be preferable.

\begin{figure}[ht]
\centering
\subfigure{
   \includegraphics[width=3.0in] {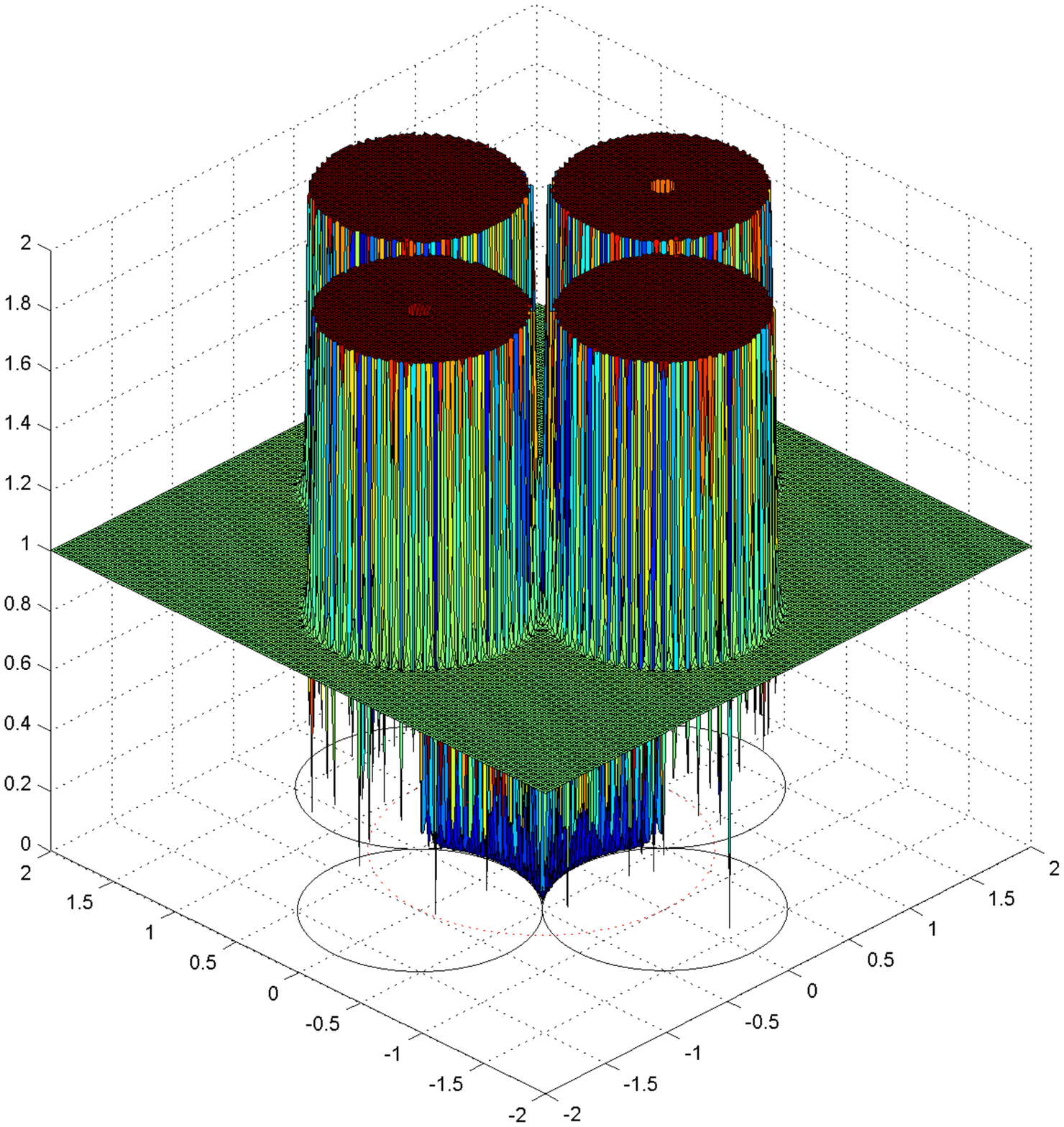}
 }
 \subfigure{
   \includegraphics[width=3.1in] {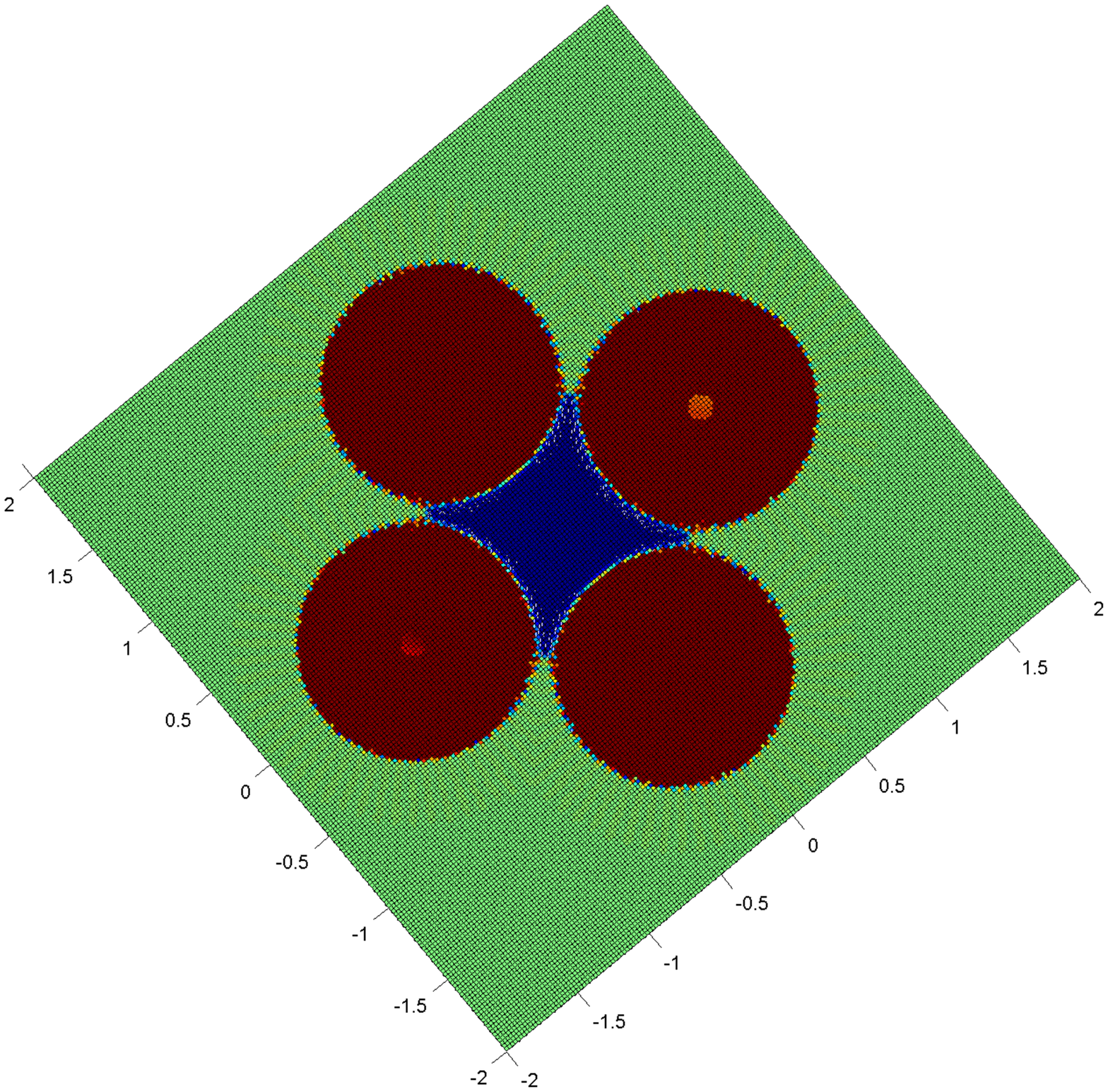}
 }
\caption{Absolute value of total pressure field  with 4 active sources, $b=1$, angle of incidence $\psi=17^\circ$,  wave number $k=2$, and $N=60$.    Values above 2 in magnitude are clipped to make the plots visible.  }
\label{figx1}
\end{figure}
\begin{figure}[h!]
\centering
\subfigure{
   \includegraphics[width=3.0in] {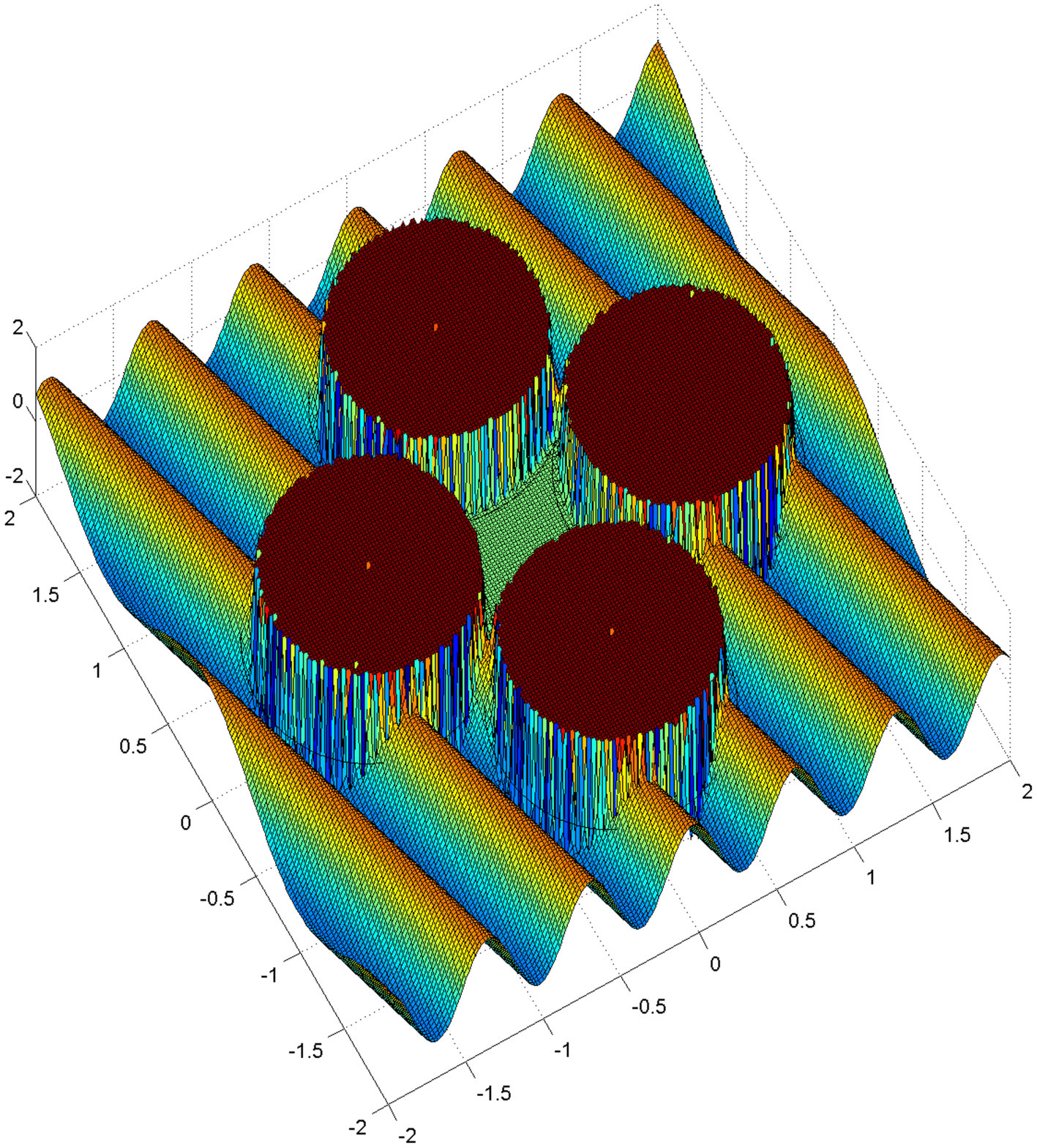}
 }
 \subfigure{
   \includegraphics[width=3.1in] {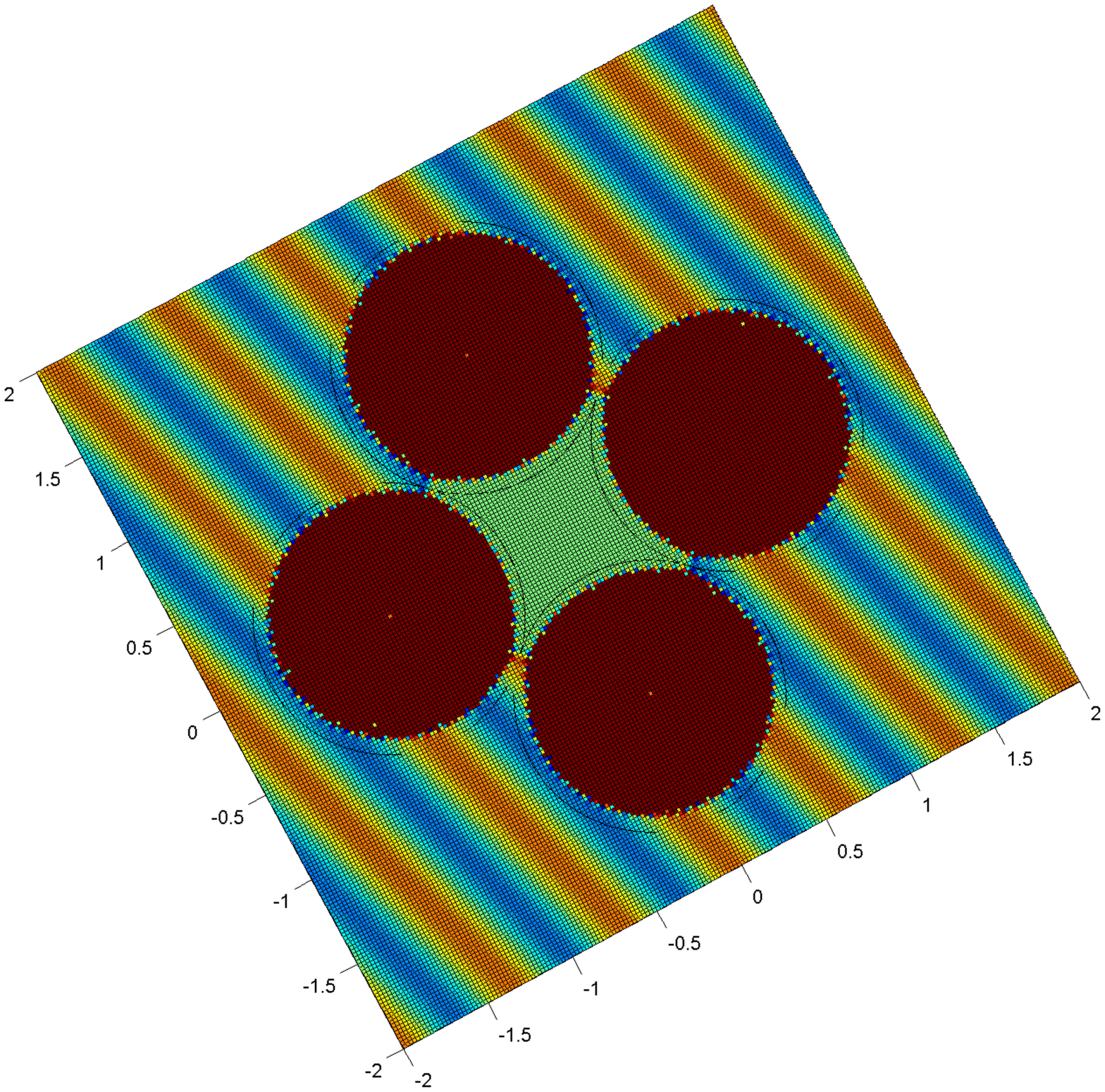}
 }
\caption{Real part of total pressure field  with 4 active sources, $b=1$,  $\psi=17^\circ$,  wave number $k=10$, and $N=60$.   }
\label{figx2}
\end{figure}

\subsection{Total field}

The total field for unit amplitude plane wave incidence on configurations of active sources of the type defined in \S\ref{Config} is illustrated through several examples.  In all cases $b=1$ and $\psi=17^\circ$.
Figure \ref{figx1} shows the absolute value of the  field for four active sources: the subplots provide different perspectives, indicating that the field is indeed essentially zero in the cloaked region $C$, and that the radiated field $u_d$ is zero outside the region $R$.    The major variation in the source field is within the circular regions centered on the active sources.   It is found that the field in these regions can take very large values, and  therefore, for the sake of visibility we truncate the plot at an arbitrary value (here = 2).   Note also that the cloaked region spills over slightly into the circular regions.  This effect is perhaps easier to see in the subsequent examples.

\begin{figure}[ht]
\centering
\subfigure{
   \includegraphics[width=3.0in] {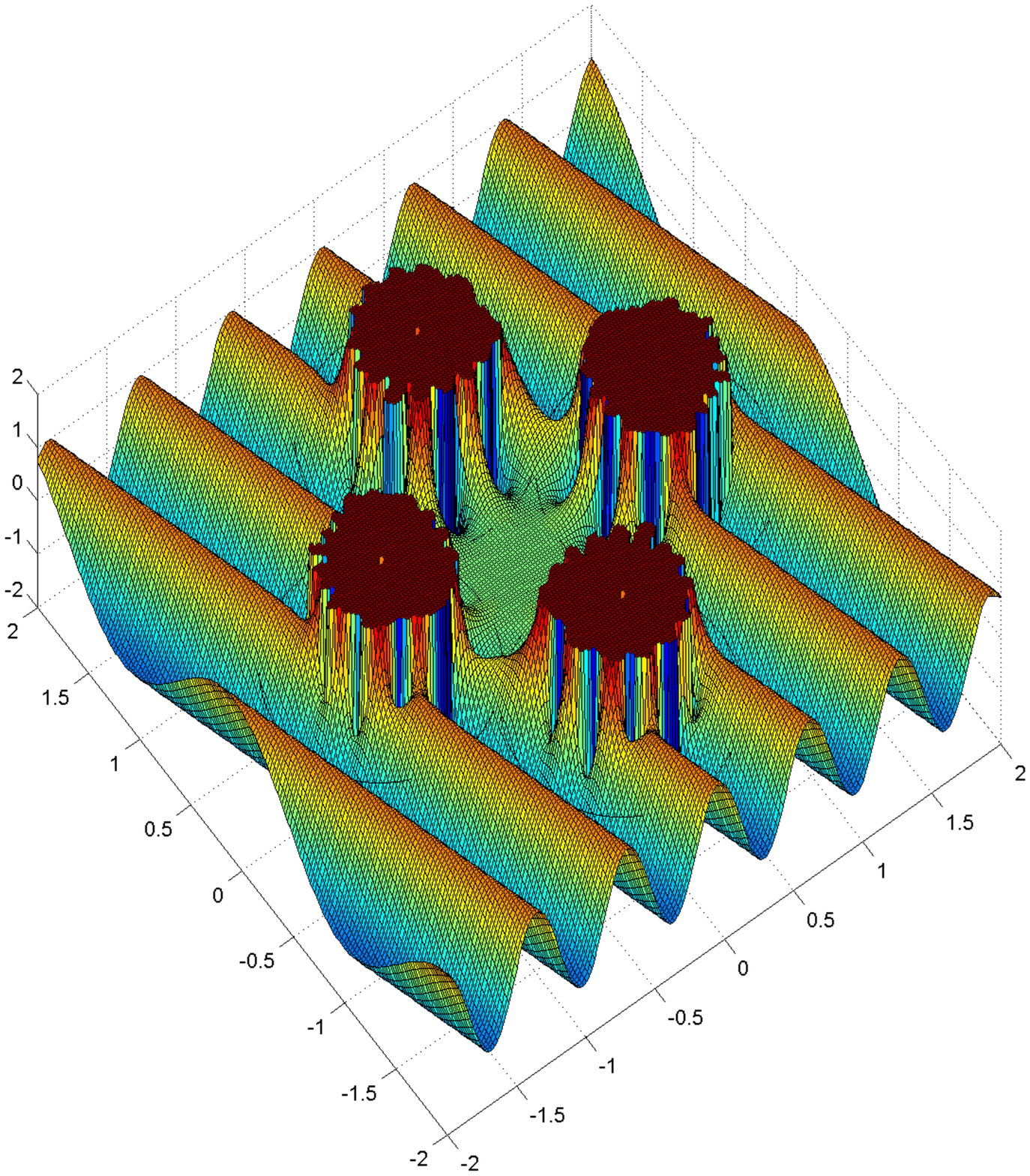}
 }
 \subfigure{
   \includegraphics[width=3.1in] {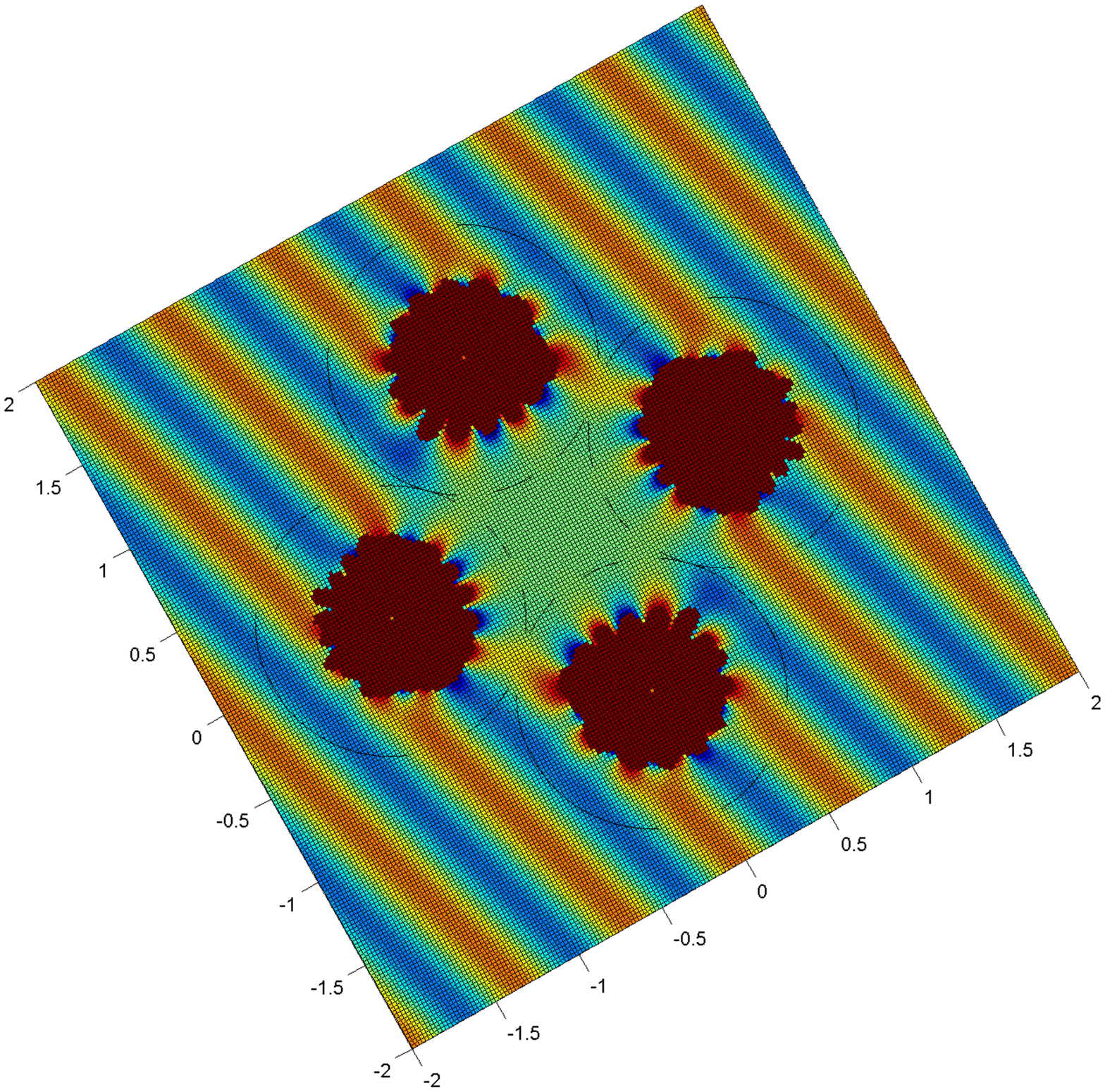}
 }
\caption{Real part of total pressure field  with 4 active sources,  $\psi=17^\circ$,  $k=10$.   The number of modes used in the truncated sum is here limited by  $N=10$.     }
\label{figx3}
\end{figure}

\begin{figure}[h!]
\centering
\subfigure{
   \includegraphics[width=3.0in] {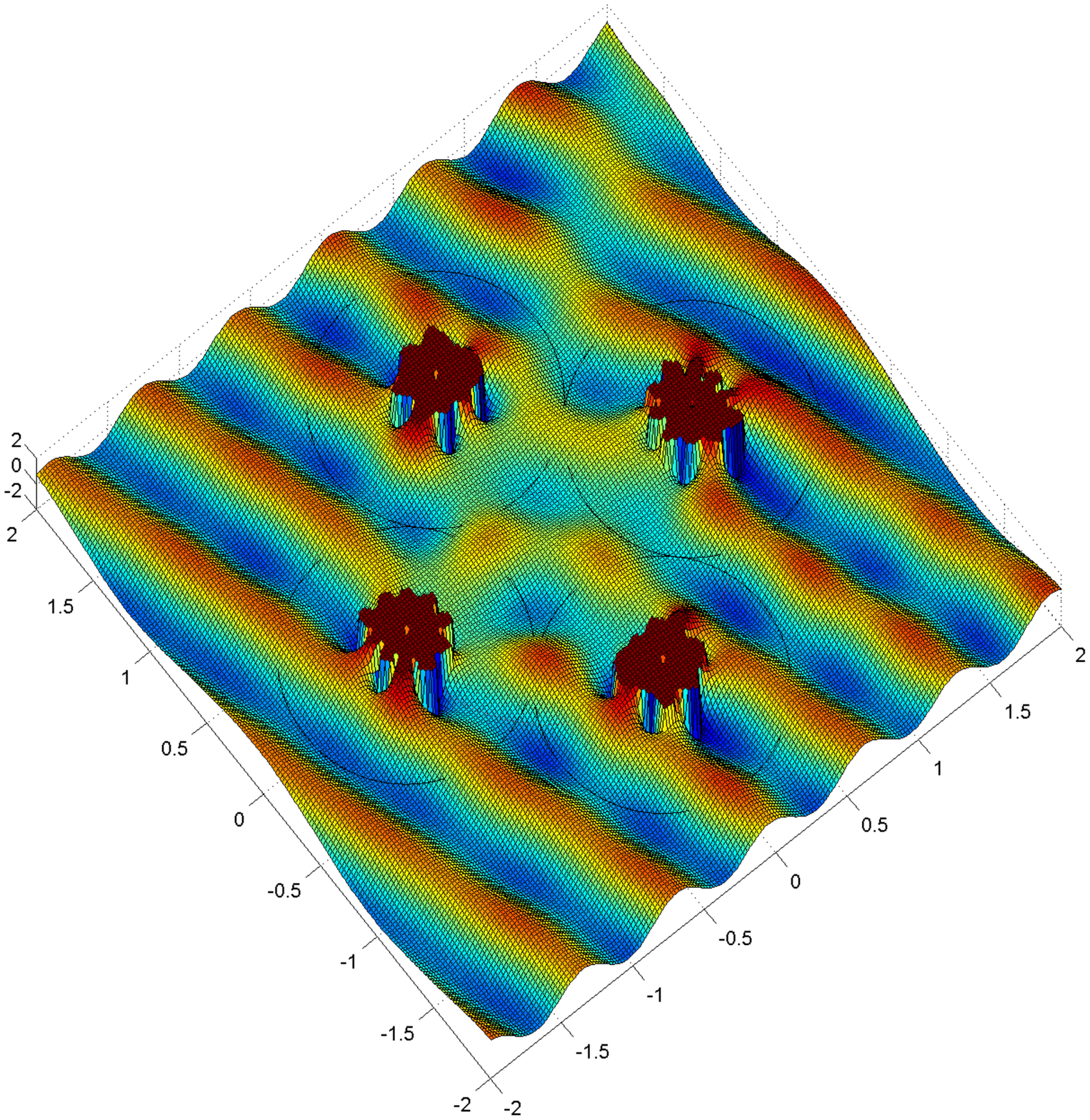}
 }
 \subfigure{
   \includegraphics[width=3.1in] {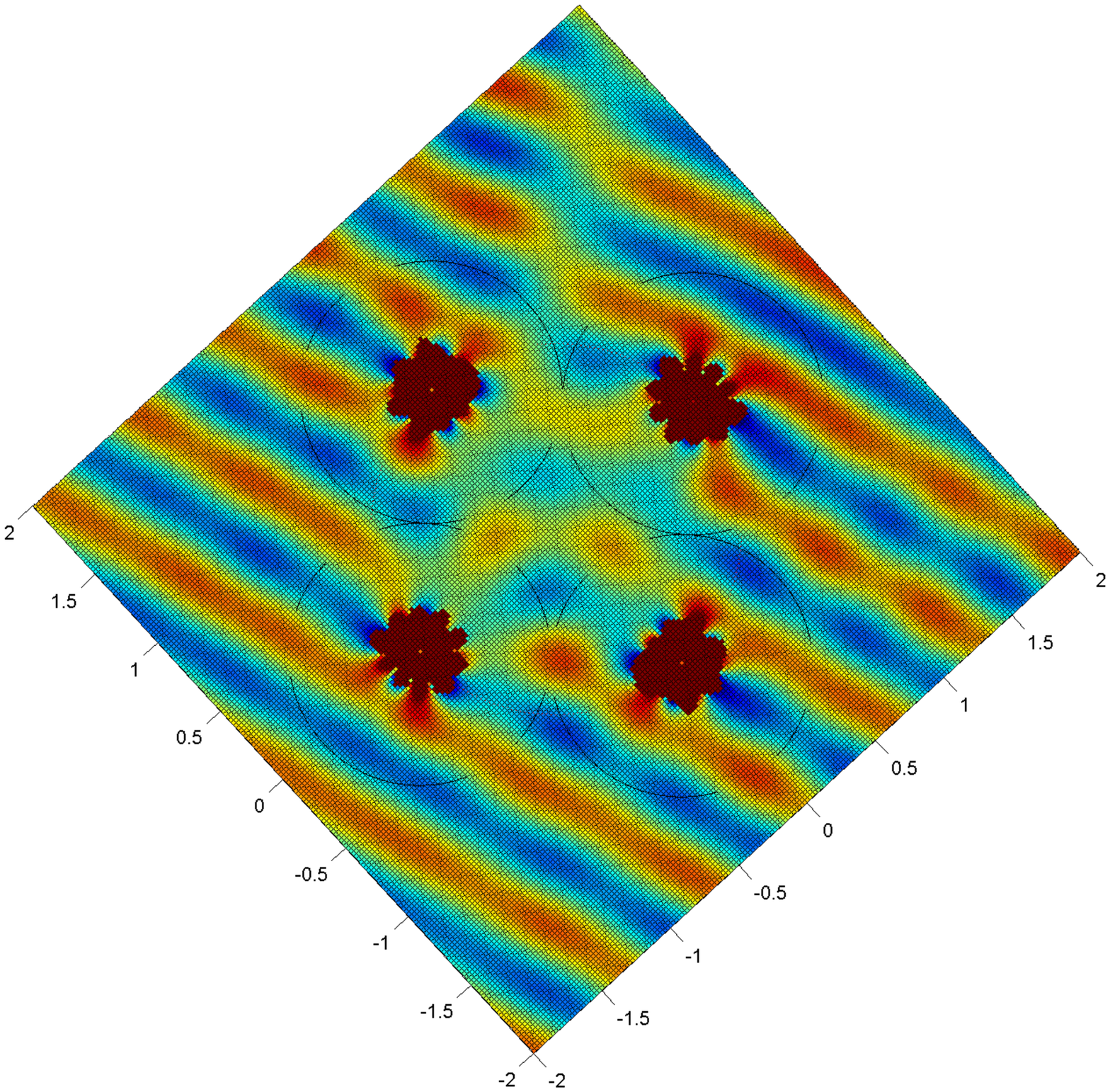}
 }
\caption{The same as in Figure  \ref{figx3} except now $N=5$.   }
\label{figx4}
\end{figure}

\begin{figure}[h!]
\centering
\subfigure{
   \includegraphics[width=3.0in] {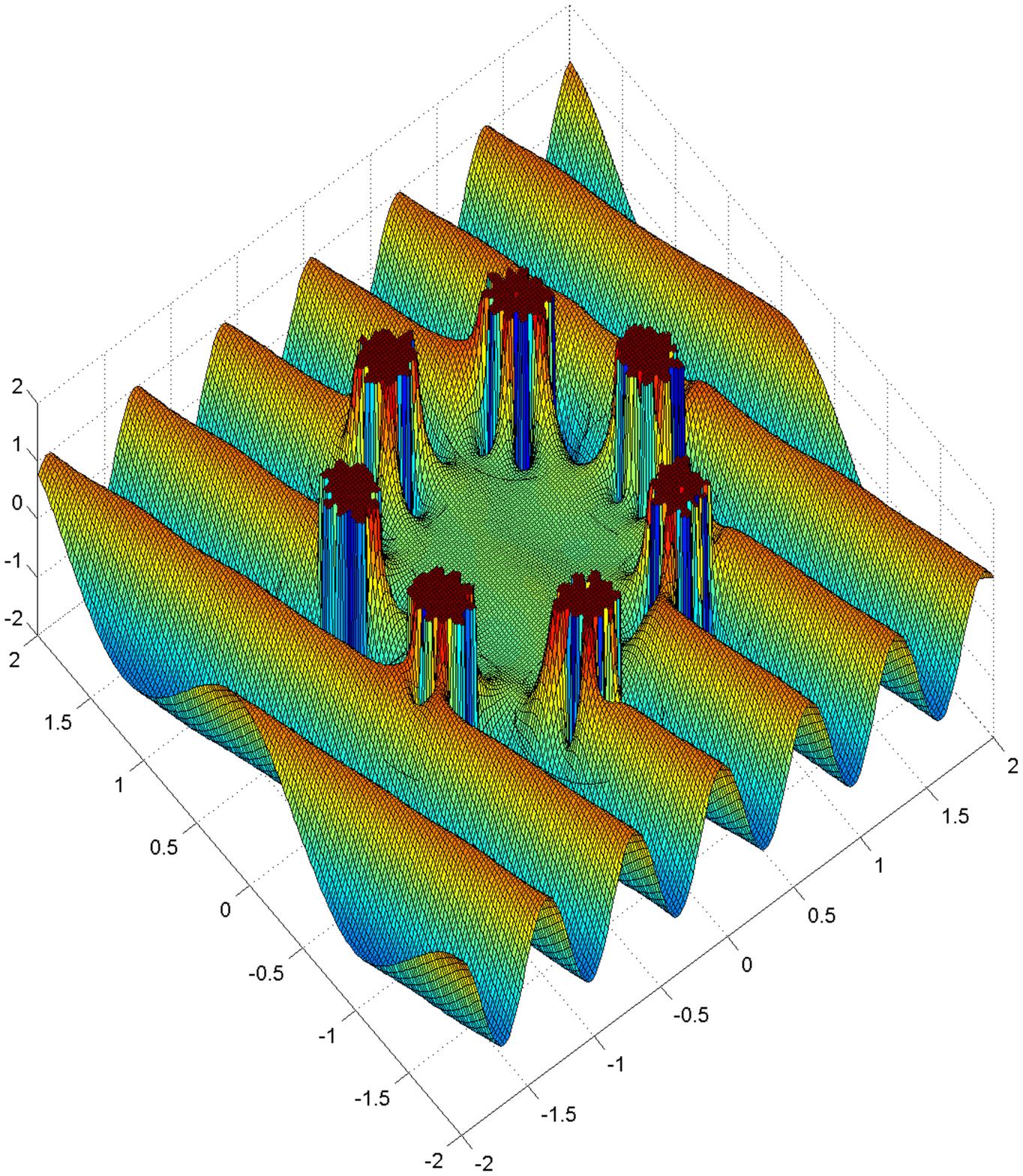}
 }
 \subfigure{
   \includegraphics[width=3.1in] {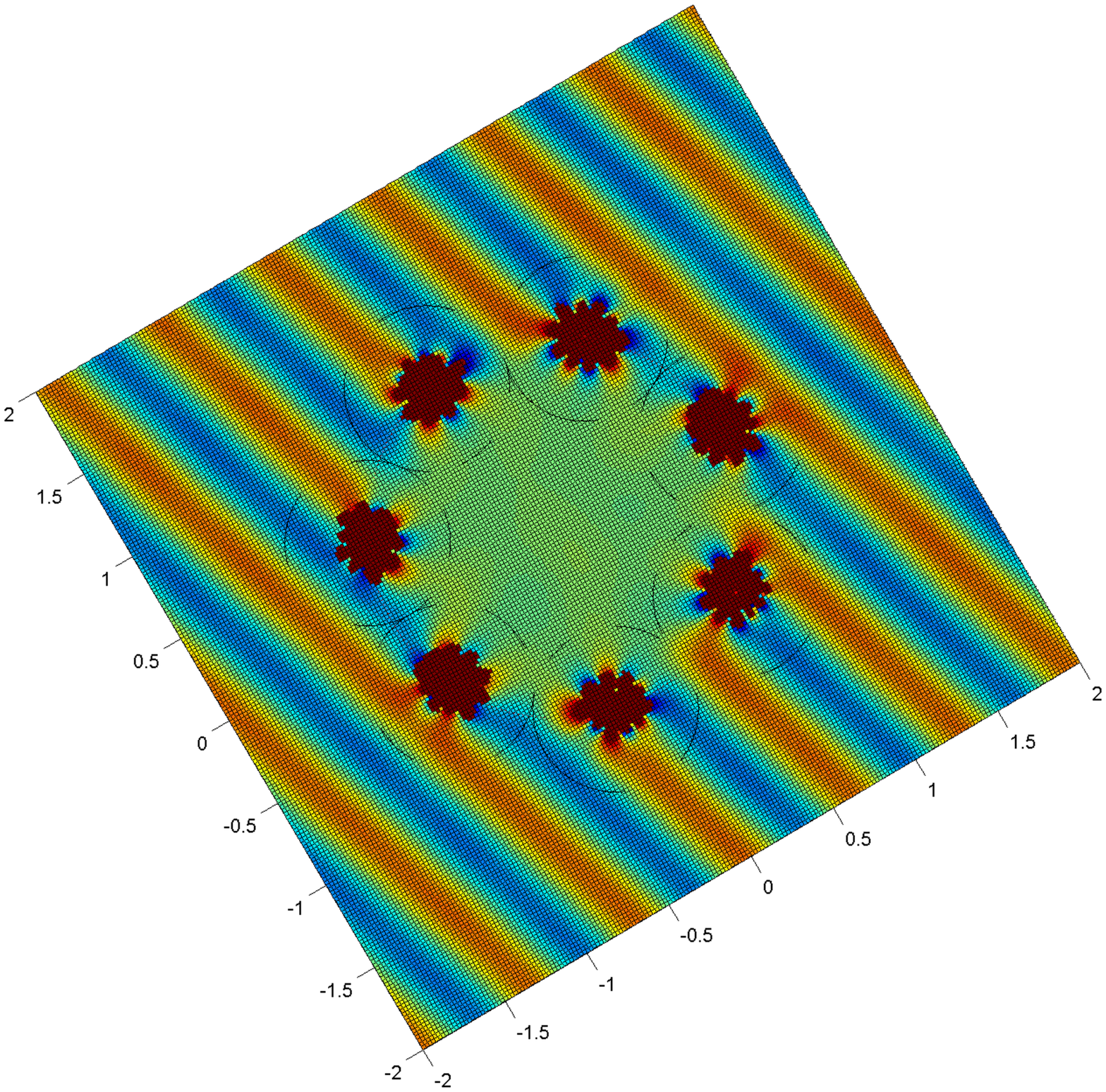}
 }
\caption{The same as in Figure  \ref{figx4} except now $M=7$.   }
\label{figx5}
\end{figure}

Figure \ref{figx2} considers the same $M=4$ configuration of active sources at a higher frequency $k=10$.  The plots  in this case show the real part of the total field, clearly illustrating the plane wave field in the exterior of $R$.  The subplot on the right clearly shows that the cloaked region is somewhat larger than $C$, extending partly into the circular regions.   The number of modes used in Figure \ref{figx2} $(N=60)$ is more than adequate to ensure convergence and accurate cloaking.  It is more instructive to consider the effect of fewer modes, as in Figures \ref{figx3} to \ref{figx5}.  In Figure \ref{figx3} the number of modes used is on the order of the frequency, and good accuracy is still observed.  Notice the smaller footprints of the active sources, as compared with Figure \ref{figx2}, indicating that the higher modes ``fill out" the regions where $u_d$ is highly variable.     Only $N=5$ modes are used in Figure \ref{figx4}, and one can see the deterioration of the cloaking effect expected with an inadequate number of multipoles.    The plane wave  is clearly evident inside the cloaked region $C$, as is some scattering effects in the ``shadow" zone.  it is interesting to note that the active source footprints are reduced in size as compared with Figure \ref{figx3}.   Finally, in Figure \ref{figx5}, we consider the effect of a larger number of active sources combined with a small number of modes.  Comparison of Figures \ref{figx4} and \ref{figx5} indicates the tendency observed from the results of \S\ref{sec42} that more active sources improves the cloaking effect. This is also to be expected from the discussion below in \S\ref{sec5} which shows that for large numbers of sources only the lowest order multipoles play a significant role.


\subsection{Scattering examples}

Finally, we  illustrate the effect of active exterior cloaking on  plane wave scattering from rigid and soft  cylinders (Neumann and Dirichlet boundary conditions, respectively).   In each case the cylinder is circular of radius $a_0=1$ centered at the origin, five active sources with $b=4$ are used, the frequency is $k =5$, and the incident wave strikes at angle   $\psi=17^\circ$.

\begin{figure}[ht]
\centering
\subfigure{
   \includegraphics[width=3.0in] {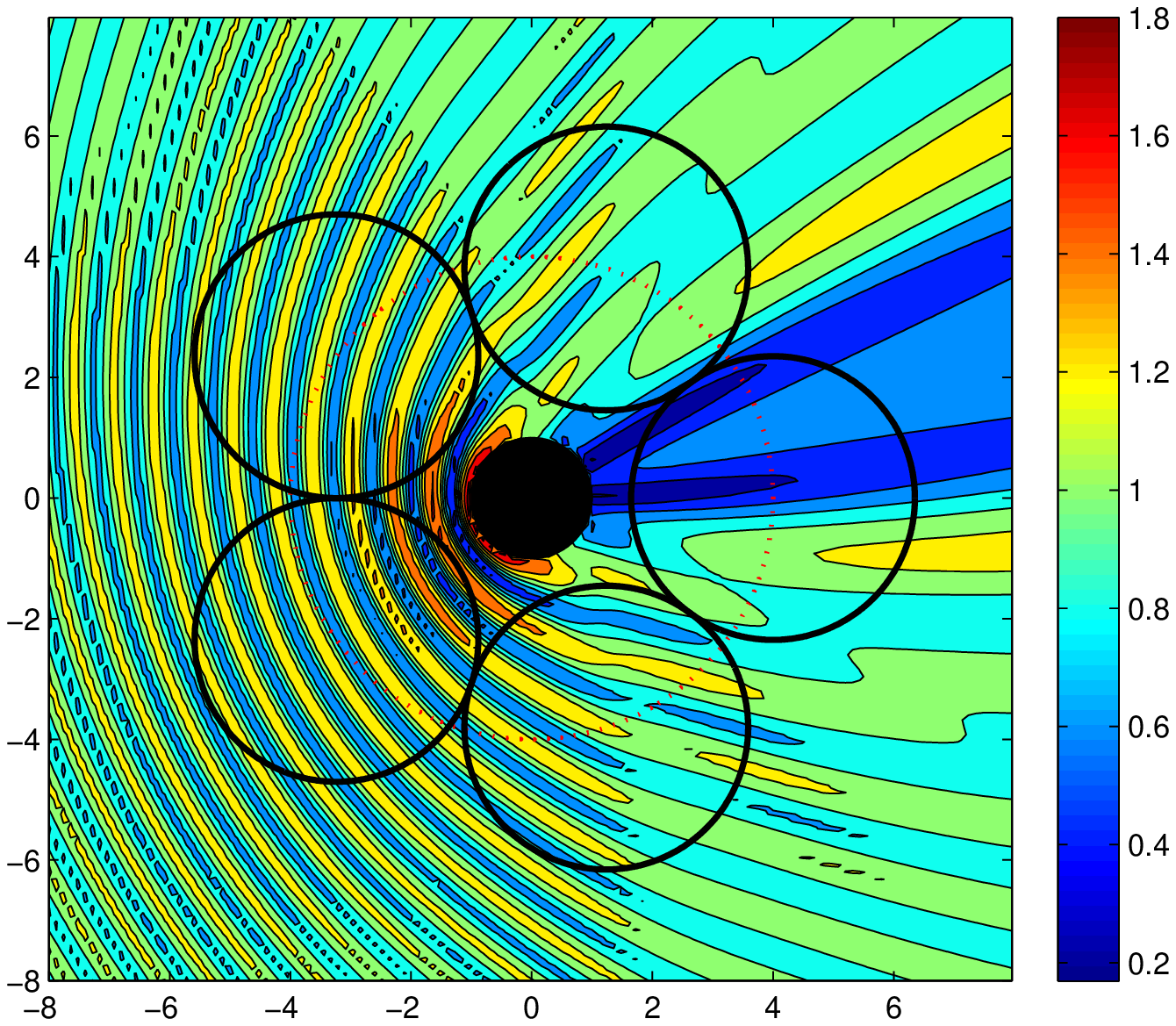}
 }
 \subfigure{
   \includegraphics[width=3.1in] {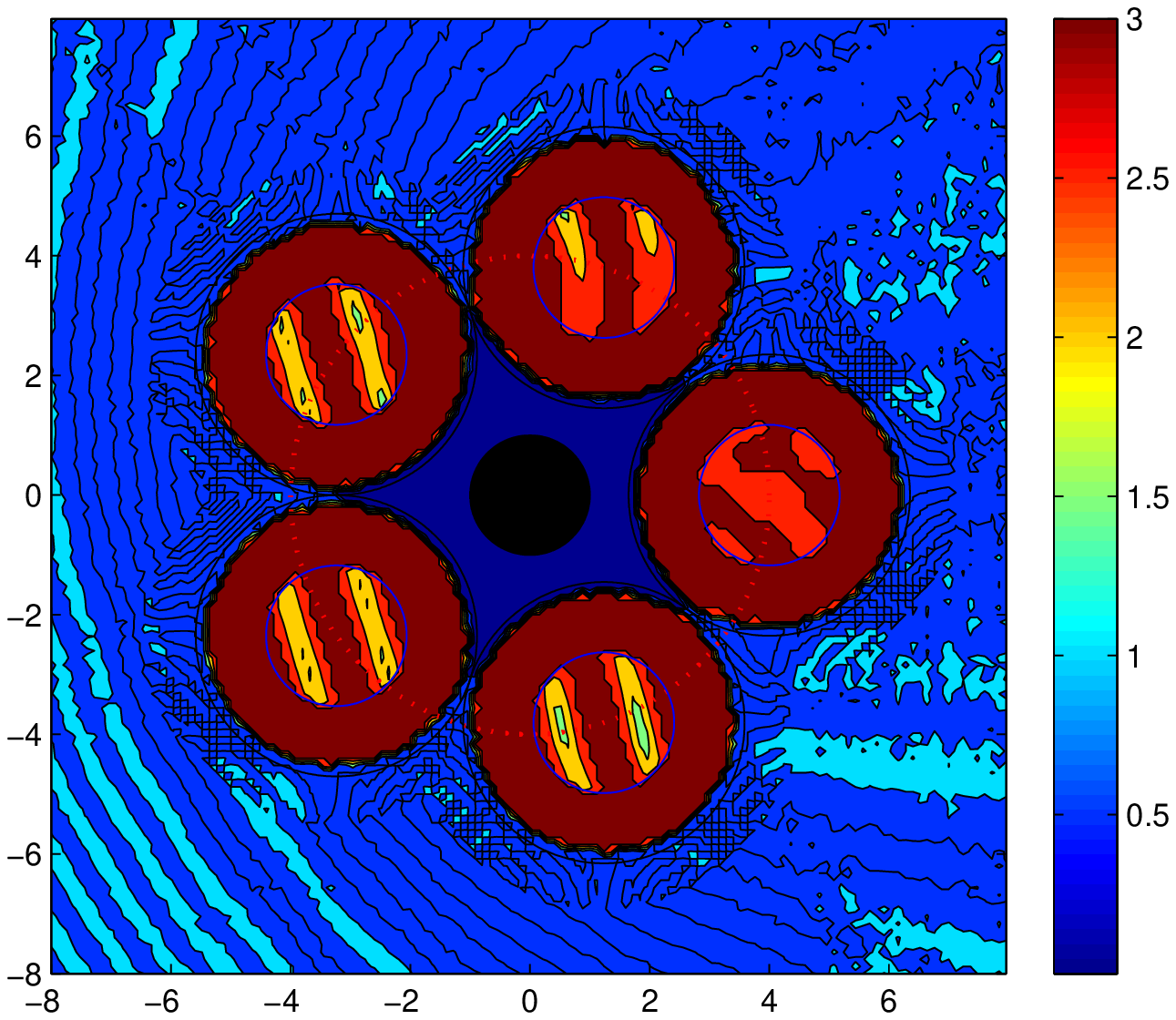}
 }
\caption{Absolute value of total pressure field  when cloaking devices are   inactive (left) and  active (right) for scattering from the hard cylinder. Calculations are performed for a hard cylinder with $M=5$ active sources, angle of incidence $\psi=17^\circ$, and wave number $ka=5$.}
\label{fig10}
\end{figure}

\begin{figure}[ht]
\centering
\subfigure{
   \includegraphics[width=3.0in] {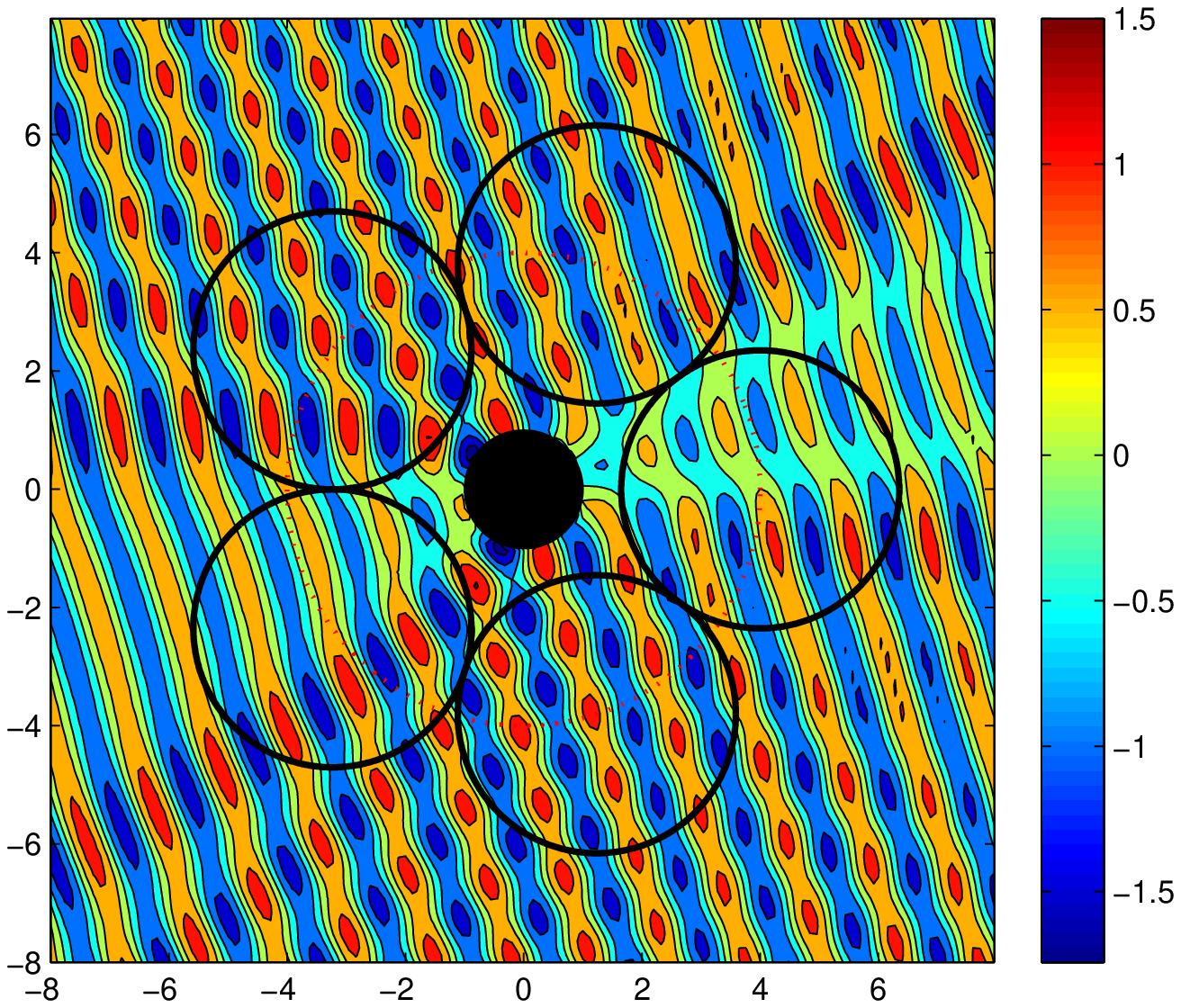}
 }
 \subfigure{
   \includegraphics[width=3.0in] {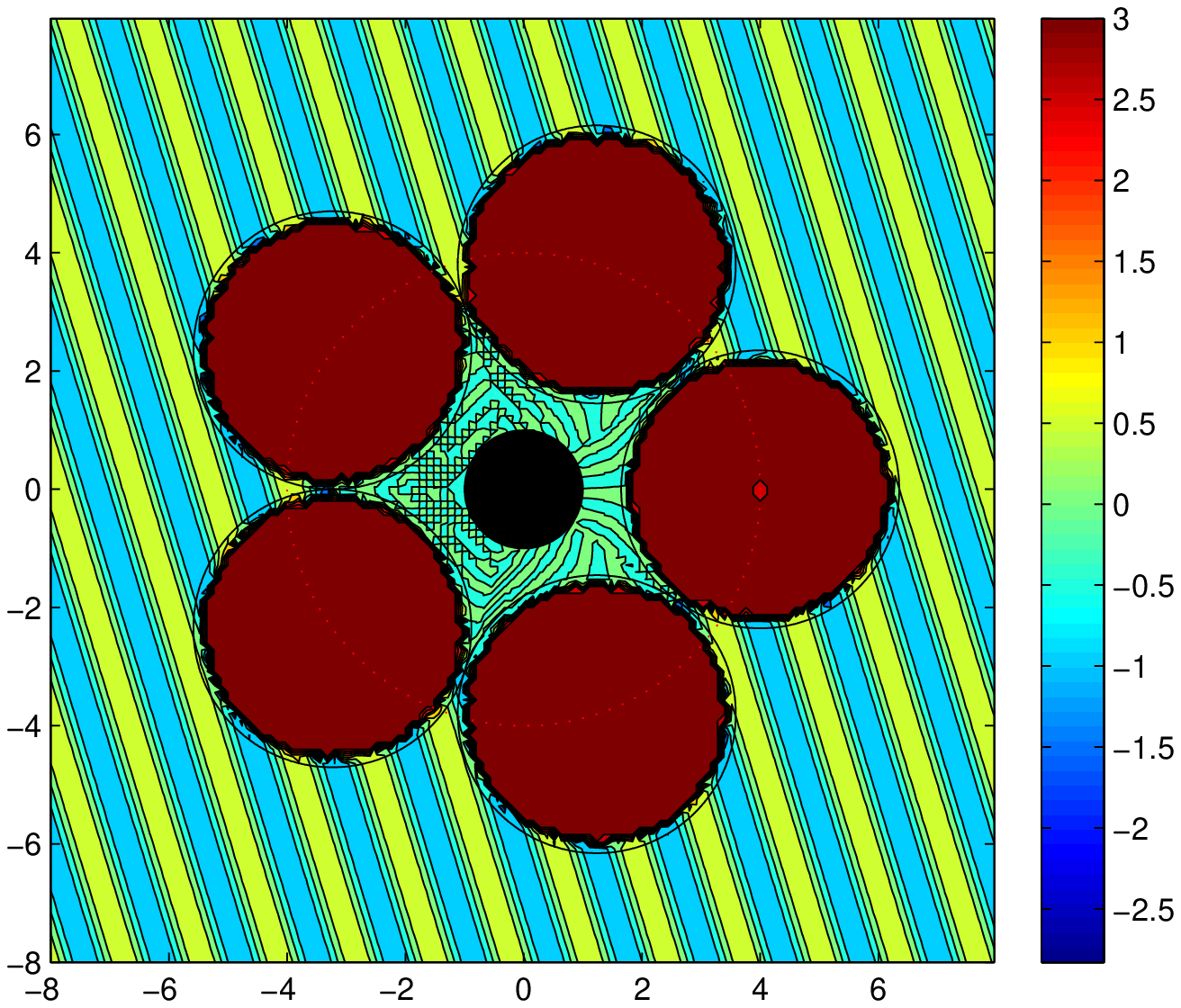}
 }
\caption{Real value of total pressure field  when cloaking devices are   inactive (left) and  active (right) for scattering from a hard cylinder.}
\label{fig11}
\end{figure}

\begin{figure}[ht]
\centering
\subfigure{
   \includegraphics[width=3.1in] {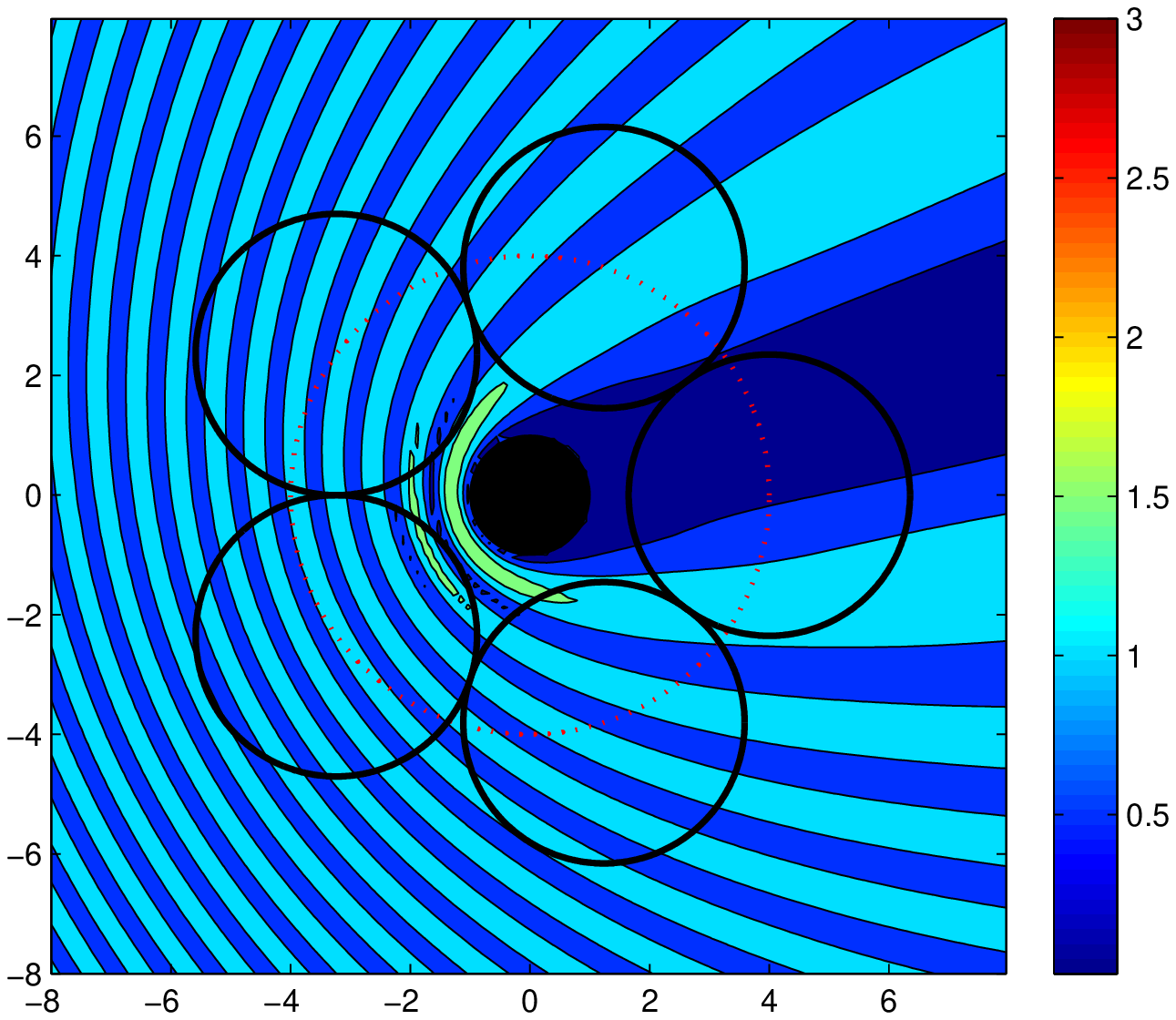}
 }
 \subfigure{
   \includegraphics[width=3.1in]  {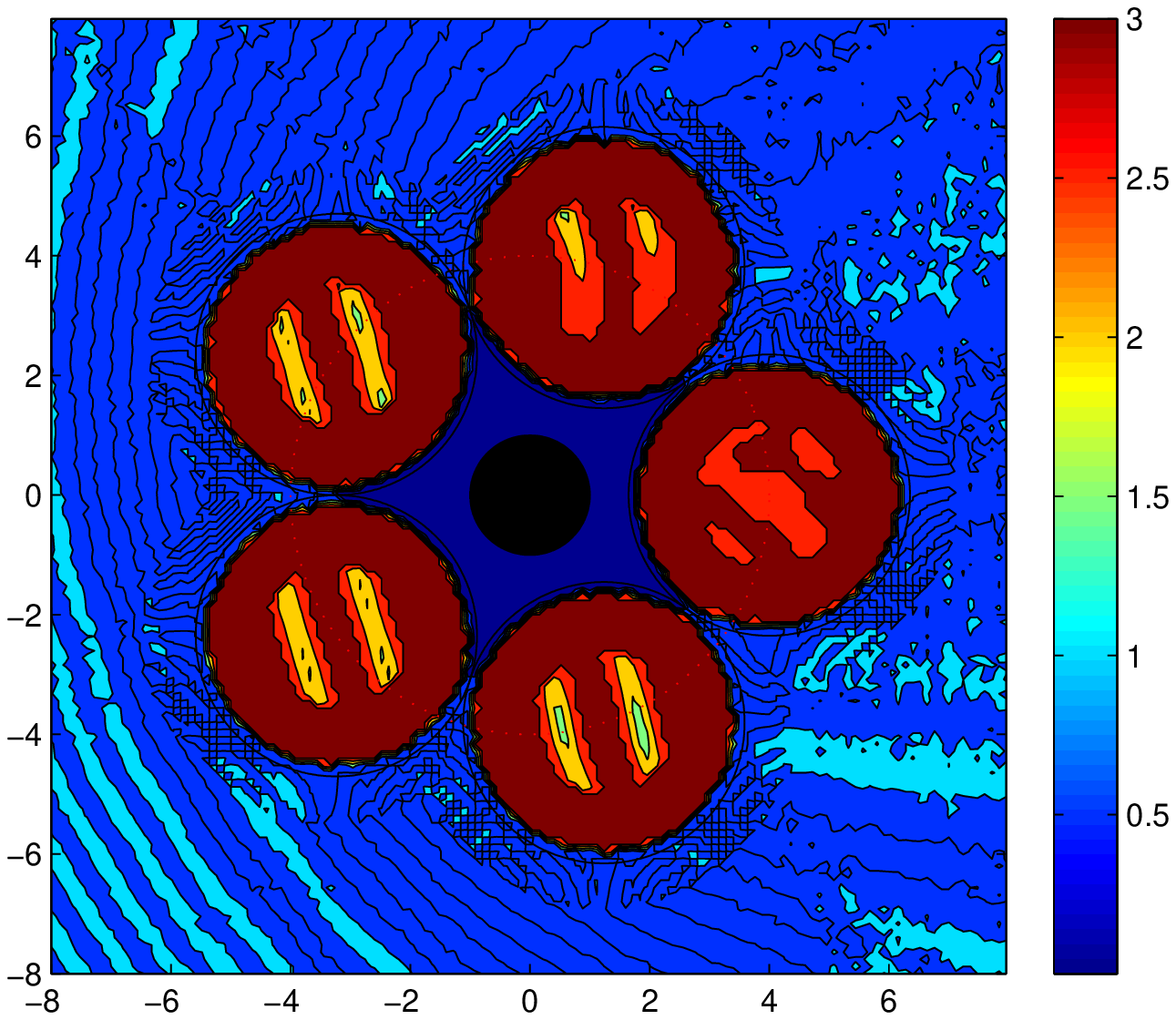}
 }
\caption{Absolute  value of total pressure field  with cloaking devices are   inactive (left) and  active (right) for scattering from a soft cylinder: $k =5$,   $\psi=17^\circ$.}
\label{fig12}
\end{figure}

Figures \ref{fig10}  and \ref{fig11} compare the response from a rigid cylinder with the active cloaking turned on  and turned off.  The absolute value is shown in  Figure \ref{fig10}  while Figure  \ref{fig11} considers only the real part of the complex field, which clearly indicates the plane wave propagating undisturbed when the cloak is active.   The comparison for a soft cylinder is shown in  Figure \ref{fig12}.

\section{Discussion} \label{sec5}

\subsection{The case of many sources:  $M \gg 1$ }
The numerical results of \S\ref{sec42} indicate better convergence properties for fewer multipoles if more active sources are used.  It is therefore of interest to consider the limit in which many sources are available: the large $M$ limit.  Staying with the configuration of \S\ref{Config}, the appropriate limit to consider is $(a_m = ) a \approx b\pi/M$ so that $ ka \ll 1$ and the small argument approximation can be used for the Bessel functions $J_n(k a_m)$  (note $kb$ is not necessarily small).   This implies that to leading order in $ka$
the coefficients in \eqref{-9b} reduce to
\beq{5--1}
b_{m,ln} = \frac i4 ka \,
\times
\begin{cases}
 (-2) \, {U_n^+}'({\bf x}_m), & l = 0,
\\
l  e^{ - il \theta_m} \,U_n^+({\bf x}_m), & l = \pm 1,
\\
0, & l \ne 0, \pm 1 .
\end{cases}
\eeq
The identity $J_{n-1}(x) - J_{n+1}(x) = 2 J_{n}'(x)$ has been used to simplify the $l=0$ term in \eqref{5--1}.
The source field    follows from eq.\ \eqref{-11b} and the identity
$V_{-1}^+  ({\bf x}) = - V_{1}^-({\bf x}_m)$ as
\beq{6-2}
\ba
u_d = & \frac i2 ka
\sum\limits_{n=- \infty }^\infty A_n
\sum\limits_{m=1}^M \times
\\ & \quad
\Big[
U_n^+({\bf x}_m) H_1^{(1)} (|{\bf x}- {\bf x}_m|)
\cos \big( \arg ({\bf x}- {\bf x}_m) - \theta_m \big)
- {U_n^+}'({\bf x}_m) H_0^{(1)} (|{\bf x}- {\bf x}_m|)
\Big].
\ea
\eeq
The  field of the active sources is therefore composed of monopoles and dipoles only, with no
contribution from  higher multipoles.  This agrees with
what one might expect from  the continuous limit of $M\to \infty$, i.e.\ a closed contour of monopoles and dipoles, but here it is obtained from the discrete solution. In fact,
eq.\ \eqref{6-2} is
\beq{6-3}
u_d = \frac i2 a
\sum\limits_{m=1}^M
\big[
u_i({\bf x}_m) \partial_n  V_0 ({\bf x}- {\bf x}_m)
- V_0 ({\bf x}- {\bf x}_m) \partial_n u_i({\bf x}_m)
\big],
\eeq
which can be seen to be the discretized version of the fundamental integral identity eq.\ \eqref{-} with the correspondence $\int \dd S \to 2a \sum_m $.
While   eq.\ \eqref{5--1} is thus the natural first approximation for $u_d$ based on the integral equation \eqref{-}, it should be realized that it was obtained here as a first order approximation of the exact expression  \eqref{-9b}.  The latter therefore provides the basis for
a multipole expansion of the exact source field obtained
 by  including higher powers of $ka$ than considered in eq.\ \eqref{5--1}.   This possibility  goes beyond our  present interests but will  be examined in a separate study dealing with approximations to the exact results of Theorem \ref{thm2}.

\subsection{Zero scattering matrix}
\ann{The exact source field $u_d({\bf x}) $ of eq.\ \eqref{-11b} exhibits some interesting features. This field is, by design, equal to the negative of the incident field in the cloaking region $\C$, and it also vanishes identically outside the concave region $R$ defined in \eqref{2=1}}
The non-radiating property of  $u_d $  is as important as the fact that it exactly cancels the incident wave in $\C$.  Let us examine this more closely.  Define
the infinite matrix ${\bf S}$ with elements $S_{pq}$ such that
\beq{-633}
 F_p = \sum\limits_{q=-\infty}^\infty S_{pq} A_q
\quad
\Rightarrow
\quad
S_{pq} =
\sum\limits_{m=1}^M\sum\limits_{l=-\infty}^\infty
b_{m,lq}   U_{p-l}^{\,-}({\bf x}_m),
\eeq
or, using \eqref{-7},
\beq{7=33}
S_{pq} =
\sum\limits_{m=1}^M \frac   {k a_m}4 \sum\limits_{l, n =-\infty}^\infty
  U_{p-l}^{\,-}({\bf x}_m)
U_{n+q}^{\,+}( {\bf x}_m )
 \frac {(-1)^{n} } {l+n}
 \left.
   \big[
   {
   {U_n^{-}}({\bf {a}}) {U_l^{-}}'({\bf {a}})- {U_n^{-}}'({\bf {a}}) {U_l^{-}}({\bf {a}})
   }\big] \right|_{{\bf a}_1^{(m)}}^{{\bf a}_2^{(m)}}.
\eeq
The matrix ${\bf S}$  is, formally at least, like a scattering matrix.  For instance,
by inspection,  ${\bf S}$ is hermitian $(S_{pq} = S_{qp}^*)$.   However, by design and based on Theorem \ref{thm1}, ${\bf S}\equiv 0$, and as such it could be called  a zero-scattering matrix.  Alternatively, it can be viewed as a formula for generating non-radiating fields.  This has relevance to the {\it inverse source problem } \cite{Tsitsas12}.  It is known that solutions to the   inverse source problem are non-unique \cite{Bleistein77}, although some   uniqueness
results are available for restricted forms of sources, e.g.
``minimum energy sources" \cite{Devaney85}.  The solution of the active cloaking problem as developed here has generated a new family of non-radiating sources, with the property that they cancel a given incident field over a finite region.

\section{Conclusions} \label{sec6}
\wjp{By definition, an active source cloaking strategy requires solution of an inverse problem: find the active source amplitudes associated with a given incident field in order to exactly cancel the latter in some finite region. The results given in Theorem \ref{thm2} provide closed-form solutions for the  inverse problem for an arbitrary time harmonic incident wave field. These new expressions require only the expansion of the incident field into entire cylindrical waves and can be evaluated to any degree of accuracy by increasing the truncation parameter $N$ associated with the number of modes of the active source. Simultaneously the fact that the active source field has been shown to vanish identically outside the region $R$ defined in \eqref{2=1} means that the active field is non-radiating. This latter property is just as important as its ability to nullify the incident wave in the region $\C$.}

\wjp{The necessary and sufficient conditions on the active source coefficients, given in Theorem \ref{thm1} provide a means to quantify the error in active cloaking when the number of modes is finite. These errors have been analyzed here in some specific scenarios. It has been shown that the error in the far-field amplitude decreases as $N$ increases, $M$ increases and $k$ decreases. In particular there is a great sensitivity to the increase in $N$; relatively small errors can be attained in the far-field amplitudes for moderate $N$, say $N\sim 10$. On the other hand for small errors in the \textit{near-field} amplitudes, relatively large values of $N$ are required. Furthermore, there is a striking reduction in error when moving from the case of $M=3$ to $M=4$ motivating the latter as a preference. In contrast to the far-field case, errors decrease for \textit{increasing} $k$.}

\wjp{Numerical results were given which illustrate the cloaking effect in various instances, including the presence of a sound-soft and sound-hard circular cylinder. In the appropriate limits, perfect theoretical active cloaking is achieved. The availability of closed-form active source amplitudes opens the door for possible studies on practical realization of active cloaking devices.}

\wjp{The case of many sources, where the active field degenerates to one involving a sum of monopole and dipole sources is worthy of further, separate study relating to the multipole expansion associated with the active field. Finally, the non-radiating nature of the active source field is especially noteworthy. The associated scattering matrix, defined in \eqref{7=33} (which is zero by design) is therefore associated with a new family of non-radiating source solutions which would appear to be useful in the so-called \textit{inverse source problem}.}


\begin{thebibliography}{10}

\bibitem{Vasquez09}
F.~G. Vasquez, G.~W. Milton, and D.~Onofrei.
\newblock Active exterior cloaking for the 2{D} {L}aplace and {H}elmholtz
  equations.
\newblock {\em Phys. Rev. Lett.}, 103:073901, 2009.

\bibitem{Miller06}
D.~A. Miller.
\newblock On perfect cloaking.
\newblock {\em Opt. Express}, 14(25):12457--12466, December 2006.

\bibitem{Vasquez09b}
F.~G. Vasquez, G.~W. Milton, and D.~Onofrei.
\newblock Broadband exterior cloaking.
\newblock {\em Opt. Express}, 17:14800--14805, 2009.

\bibitem{Vasquez11a}
F.~G. Vasquez, G.~W. Milton, D.~Onofrei, and P.~Seppecher.
\newblock {Transformation elastodynamics and active exterior acoustic
  cloaking}.
\newblock In S.~Guenneau and R.~Craster, editors, {\em Acoustic Metamaterials:
  Negative Refraction, Imaging, Lensing and Cloaking}, pages 1--1. Canopus
  Academic Publishing and Springer SBM, 2012.

\bibitem{Vasquez11}
F.~G. Vasquez, G.~W. Milton, and D.~Onofrei.
\newblock Exterior cloaking with active sources in two dimensional acoustics.
\newblock {\em Wave Motion}, 49:515--�524, 2011.

\bibitem{Onofrei11}
D.~Onofrei and K.~Ren.
\newblock {On the active manipulation of quasistatic fields and its
  applications}.
\newblock {\em {arxiv.org/abs/1109.4182}}, 2011.

\bibitem{Vasquez11b}
F.~G. Vasquez, G.~W. Milton, and D.~Onofrei.
\newblock {Mathematical analysis of the two dimensional active exterior
  cloaking in the quasistatic regime}.
\newblock {\em {arxiv.org/abs/1109.3526}}, 2011.

\bibitem{Milton06b}
G.W. Milton and N.A.P. Nicorovici.
\newblock On the cloaking effects associated with anomalous localized
  resonance.
\newblock {\em Proc. R. Soc. A}, 462:3027�3059, 2006.

\bibitem{Zheng10}
H.~H. Zheng, J.~J. Xiao, Y.~Lai, and C.~T. Chan.
\newblock {Exterior optical cloaking and illusions by using active sources: A
  boundary element perspective}.
\newblock {\em Phys. Rev. B}, 81:195116+, 2010.

\bibitem{Tsitsas12}
N.~L. Tsitsas and P.~A. Martin.
\newblock {Finding a source inside a sphere}.
\newblock {\em Inverse Problems}, 28(1):015003+, January 2012.

\bibitem{Devaney85}
A.~J. Devaney and R.~P. Porter.
\newblock {Holography and the inverse source problem. Part II: Inhomogeneous
  media}.
\newblock {\em J. Opt. Soc. Am. A}, 2(11):2006--2011, 1985.

\bibitem{Bleistein77}
N.~Bleistein and J.~K. Cohen.
\newblock {Nonuniqueness in the inverse source problem in acoustics and
  electromagnetics}.
\newblock {\em J. Math. Phys.}, 18(2):194--201, 1977.

\bibitem{Abramowitz74}
M.~Abramowitz and I.~Stegun.
\newblock {\em Handbook of Mathematical Functions with Formulas, Graphs, and
  Mathematical Tables}.
\newblock Dover, New York, 1974.

\bibitem{Colton}
D.~L. Colton and R.~Kress.
\newblock {\em Integral Equation Methods in Scattering Theory}.
\newblock Krieger, Melbourne, FL, 1991.

\end{thebibliography}


\end{document}